# AAO OBSERVER

# Multi-Object IFU comes to the AAT

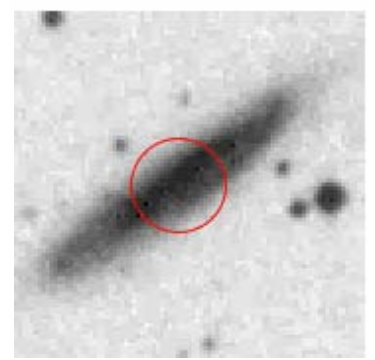

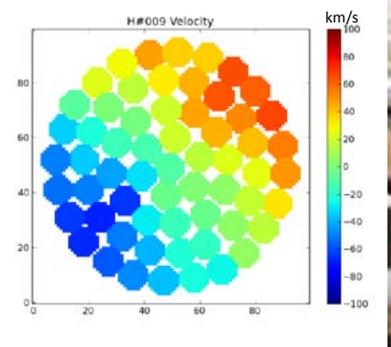

The 6dFGS Fundamental Plane | Are you biased? | Dragonfly flutters its wings

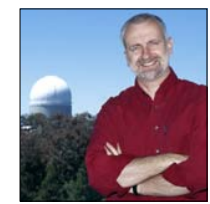

# Director's message
Matthew Colless

We are currently in the process of developing the AAO *Forward Look*, a strategic plan that will define the AAO's goals for 2011-2015 and beyond. It is based on the goals and priorities set out in the *Australian Astronomy Decadal Plan 2006-2015* and the recent *Mid-Term Review of the Decadal Plan* (see http://www.science.org.au/natcoms/nc-astronomy/decadalplan.html). The AAO Advisory Committee endorsed the *Forward Look* process at its inaugural meeting in March, and initial consultations have already been held with the Advisory Committee, the AAL Board, AAL's Optical Telescopes Advisory Committee, the AAO Users' Committee, the Australian Time Assignment Committee and AAO staff.

There are many challenges and opportunities for the AAO in the next few years, but the eight framing goals for the *Forward Look* are:

**1. Maximising the research productivity and impact of both the AAO itself and the users of its facilities.** This is the fundamental goal of the AAO, and sets the context in which all other issues are addressed. The AAO aims to be a world-class astronomical institution, providing excellent optical and infrared observing facilities and innovative telescope instrumentation that enable Australian astronomers to do outstanding science (as evidenced by the articles on pages 6, 10, 11 and 15 of this issue).

**2. Determining the effective scientific lifetimes of the AAT and UKST, and developing appropriate and cost-effective operations models.** World-class research requires an appropriate mix of facilities on all scales. In the next 5 years the *Decadal Plan* and *Mid-Term Review* recommend that Australia increase its access to large optical facilities to at least the equivalent of a 20% share in an 8-metre telescope. On a ten-year timescale, Australian astronomers aim for a 10% share in an 'Extremely Large Telescope', an ambition currently realised by Australia's participation in the 25-metre Giant Magellan Telescope project. The effective lifetimes of the AAT and UKST depend on their scientific competitiveness with respect to such facilities, which in turn depends on telescope capabilities, instrumentation suites, levels of access, scientific agendas, and the operational funding available to support Australia's portfolio of optical telescopes. With the ongoing refurbishment program, new instruments such as HERMES, and upgrades to the existing instrument suite, the AAT will remain scientifically competitive for another decade while also remaining a valuable testbed for new instruments and technologies. Over this period, however, the AAO, which already supports Australian access to Gemini and Magellan, will shift its operational emphasis towards these larger telescopes and GMT. In the meantime, the UKST can continue operating in user-pays mode, particularly if refurbishment of the telescope and an upgrade to 6dF allow more ambitious programs, such as the proposed TAIPAN galaxy survey.

**3. Managing the AAO's evolving role at Siding Spring Observatory in light of foreshadowed changes in ANU's role and support.** Over the next five to ten years the ANU is likely to be scaling back its level of support for operations at Siding Spring Observatory (SSO). Appropriate evolution of the operations model for SSO therefore needs to be considered, including the possibility that the AAO might assume responsibility for SSO operations. In that case the ANU would continue to own the site, but become one of several organisations with facilities at SSO that are supported by the AAO.

**4. Improving the AAO's support model for offshore telescopes where Australia is a partner in an international consortium.** At present the AAO's focus is on operating its onshore telescopes (AAT & UKST), but it also supports Australian users of offshore telescopes (Gemini & Magellan). As the focus and investments shift to larger offshore telescopes, the AAO must develop a plan for maximising outcomes from such facilities. This will involve some appropriate mix of organisational support (for time allocation committees, user committees and so on), user support (for preparation of proposals, remote observing capabilities, expert assistance in data reduction and the like) and instrumentation development (setting scientific agendas and winning guaranteed time for the community). Together these services must provide evident added value to Australian users of these facilities.

**5. Planning the AAO's next generation of instruments for all these telescopes, and leveraging the best scientific opportunities for Australian astronomers through the instrumentation program.** The AAO has one of the world's best astronomical instrumentation programs and is a leader in wide-field multi-object spectrographs and robotic fibre positioners. The AAO also has a unique and innovative instrument science group developing new technologies such as OH-suppression fibres, hexabundles, starbug robots and integrated photonic spectrographs (see the articles on pages 4, 7 and 13 of this issue). These capabilities are already exploited to keep the AAT competitive and to leverage additional access to front-rank facilities by providing instruments. The latter approach will become increasingly important for the AAO and Australian astronomers as the importance of Australia's onshore facilities declines relative to international offshore facilities. The next generation of AAO instruments needs to be matched to the scientific lifetimes and operational models of the AAT and UKST, and balanced with the opportunities to gain additional access to international facilities in which Australian astronomers have an interest.

**6. Exploiting the improved facilities of the AAO's new Sydney headquarters to energise and advertise the organisation.** The move of the AAO's headquarters to new premises in North Ryde, slated for the middle of next year, represents a significant investment in the organisation by the Australian government. The new building will provide improved office areas and better facilities for the instrumentation program, allowing the AAO to efficiently assemble, integrate and test larger instruments for larger telescopes. Together with the solid funding outlook for the Observatory, the new headquarters will provide the AAO with the confidence and capacity to make the changes needed to respond to the changing astronomical environment and to undertake more ambitious instrument projects. This in turn will energise AAO staff and advertise to the world that the AAO continues to be a force in international astronomy.

**7. Recruiting and nurturing world-class staff.** A research institution's staff is its most important resource, so it is key to the AAO's future that it continues to recruit and nurture world-class astronomers, instrument scientists and engineers. This will be possible if the AAO is recognised internationally as a powerhouse of astronomical research and technology development, and if it provides facilities and opportunities that excite and challenge the best young people in Australia and from overseas.

**8. Maintaining good relations with the astronomy community by being responsive to changing needs and effective in delivering services.** The AAO's success and strong support in the Australian astronomical community is founded upon its track record of responsiveness to community needs and effectiveness in providing competitive facilities and services for researchers. Close consultation with the community is essential to maintain this situation in future, and the *AAO Forward Look* must therefore be consulted with, and owned by, the community if it is to achieve its goals.

We plan to make a consultation draft of the *Forward Look* available to the Australian astronomical community in early November, with the final version to be completed and made public by the end of the year. In order to ensure that the community has every opportunity to discuss the draft and provide feedback, I will be visiting major centres to hold town-hall meetings discussing the *Forward Look* during November and early December. The AAO's *Forward Look* will be critical to the successful evolution of the AAO as Australia's national optical observatory over the next five to ten years, so I look forward to these discussions!

# CONTENTS

# SAMI Commissioning – First Hexabundle Galaxy Spectra

Jon Lawrence (AAO), Scott Croom (University of Sydney), Amanda Bauer (AAO), Joss Bland-Hawthorn (University of Sydney), Sarah Brough (AAO), Julia Bryant (University of Sydney), Matthew Colless (AAO), Simon Ellis (AAO), Tony Farrell (AAO), Lisa Fogarty (University of Sydney), Michael Goodwin (AAO), Ron Heald (AAO), Anthony Horton (AAO), Andrew Hopkins (AAO), Heath Jones (Monash/AAO), Steve Lee (AAO), Geraint Lewis (University of Sydney), Stan Miziarski (AAO), Samuel Richards (AAO/University of Sydney), Max Spolaor (AAO)

Hexabundles are a recently developed technology (Bryant et al, 2011) that offer the potential for integral-field spatially-resolved spectroscopy without the complexities inherent with existing bulk-optic lenslet array, microlens array, or image-slicing mirror techniques. The hexabundle consists of a series of multimode fibre cores (so far 7, 19, and 61 core devices have been demonstrated) that are lightly fused together over a small (~20 mm) interaction length. Interstitial holes are filled with soft, low refractive index glue. This geometry acts as a compact relatively high fill-factor integral-field unit (IFU)/image-slicer, giving significant advantages for multi-object IFU systems that bridge the gap between large monolithic IFUs and single-aperture multi-object systems.

## The Hexabundle Advantage

The key benefits of integral-field spectroscopy for extra-galactic science are that gas and stellar kinematics over an entire galaxy can be measured enabling the separation of dynamical components, the measurement of dynamical mass, the examination of the impact of winds and out flows, and the discovery of merging systems via dynamical disturbance. This information is obtained in addition to parameters than can be readily measured with single-aperture spectroscopic instruments (e.g., star formation rates, gas phase metallicities, stellar ages, stellar metallicities, and black hole accretion and extinction due to dust).

Integral field spectroscopy has so far almost exclusively been limited to single-object instruments, meaning that it is time consuming to build large galaxy-samples. The key innovation of the hexabundle-based approach is that it becomes possible to build systems that are capable of positioning a significant number (i.e., tens to hundreds) of IFUs simultaneously. If fed at a low enough focal ratio, optics are not required in the fibre bundle, significantly simplifying alignment and assembly. Additionally, each hexabundle IFU can be made much smaller than its bulk- or micro-optics counterpart – simplifying the positioner system design and allowing close packing of objects.

## The SAMI Instrument

With a view to providing the first on-telescope demonstration of hexabundle technology, the AAO and the University of Sydney have collaborated on the development of the SAMI instrument for the AAT. SAMI uses 13 x 61 core unfused hexabundles that are mounted on a plug-plate at the 1 degree field-of-view triplet corrector top-end focus (Figure 1) on the Prime Focus Camera. At f/3.4 with 105 µm core diameter fibres, each hexabundle samples a 14" field at 1.6" per fibre. At the output end, a total of 13 V-groove slit blocks are mounted at the AAOmega slit. Each slit block includes 63 fibres (all the fibres from 1 hexabundle and 2 fibres for sky subtraction).

A ribbonised fibre cable of length ~40m joins the two instrument ends. Fibres from this cable are glued into the V-grooves at the spectrograph slit. At the top end, the hexabundle fibres are each fusion spliced to the ribbon fibres inside a protective "splice box", mounted on the internal wall of the top-end barrel.

Each hexabundle is mounted in a standard (SMA) screw-thread fibre connector. The plug plates have a mating connector at each object position. Two galaxy fields (i.e., 26 objects) are pre-drilled per plate along with a set of 26 blank sky locations common to both galaxy fields. For the proposed integration time of 2 hours per field this means that 3 plates (or 2 plate exchanges) are required each observing night. The down time between fields is less than 30 minutes.

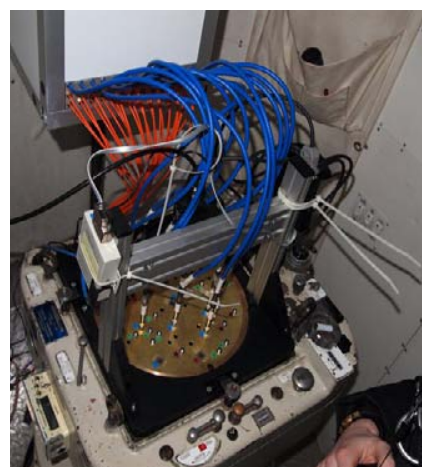

Figure 1. The SAMI plug plate assembly unit mounted onto the Prime Focus Camera. The white "splice box" connects the blue hexabundles and orange sky fibres from the brass plug-plate to the fibre bundle.

For acquisition and guiding we use a CCD camera mounted on a gantry above the plug plate that views the central few arcmins of the field through a hole in the plate.

## SAMI Commissioning

The first commissioning run for the SAMI multi-object IFU instrument occurred from 1st-4th July 2011. After installing the instrument, the initial step was to align the plug plate to the telescope optical axis. It was unsure prior to SAMI commissioning whether the Prime Focus Camera that was originally built for commissioning of the AAT in the early 1970s and last used over a decade ago, would be well aligned to the telescope optical axis. This provided the opportunity/necessity for an observer to ride in the top end (Figure 2 and front page), something also not seen at the AAT for some years.

Initial coarse alignment of the plug plate assembly was done by eye – using holes positioned around several bright (m~6) stars in the field. Fine adjustment was accomplished using fainter (m~14) stars coupled through the hexabundles and the AAOmega spectrograph. A near real-time data pipeline extracted the partially-reduced spectra from the detector images and incorporated a schematic display of reconstructed hexabundle fields based on total integrated counts per hexabundle core (Figure 3). Measured centroids were used to derive precise offsets for each bundle – allowing field rotation and telescope pointing corrections. These data were also used to derive corrections to the field distortion coefficients that will be used for the next commissioning run.

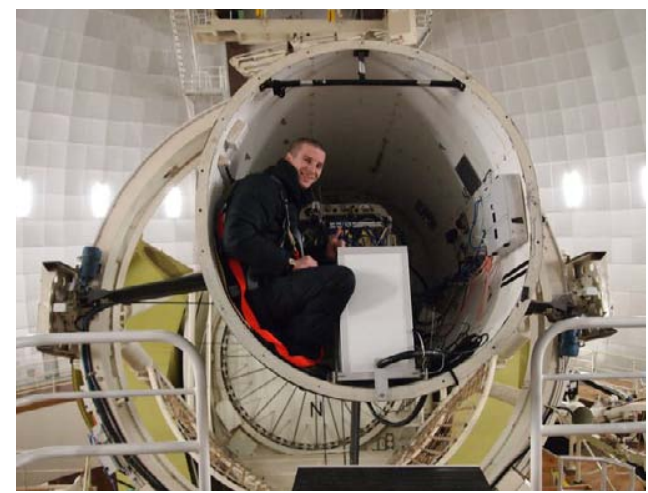

Figure 2. Team member Sam Richards prepares for a long cold night in the triplet top end.

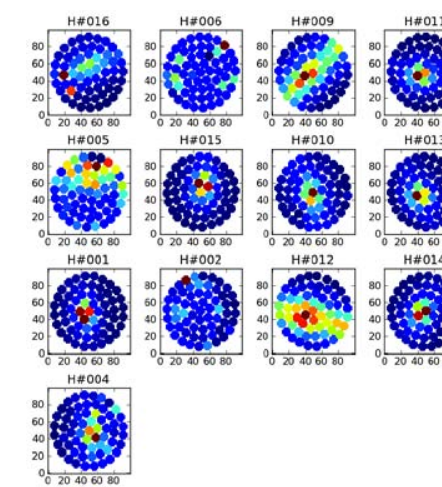

Figure 3. Schematic display of reconstructed hexabundle images based on total integrated counts per hexabundle core for a field with 13 galaxies. Note that hexabundles #6 and #2 were not functional during this observation. Galaxy positions were all centred within ~2 cores.

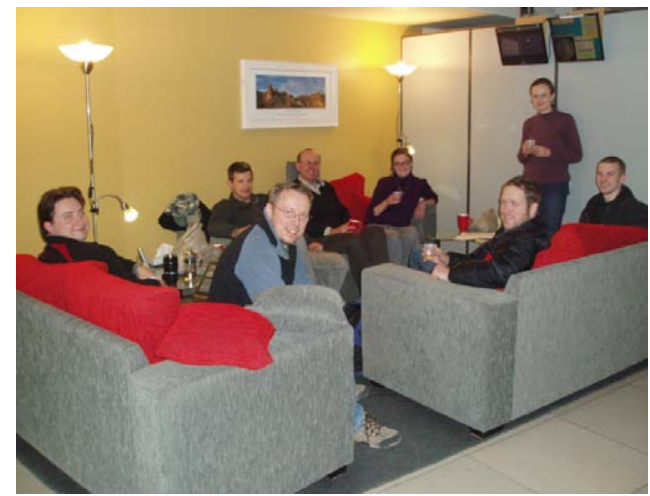

Figure 4. SAMI commissioning team galaxy-field celebrations.

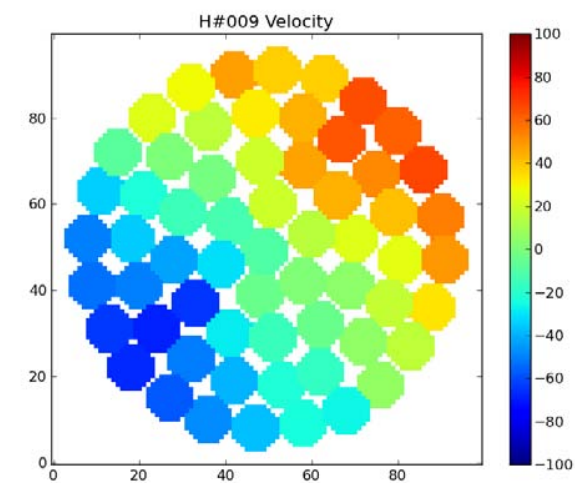

Figure 5. Velocity map produced from the first galaxy-field commissioning plate. The box is ~15" square.

Early in the morning on the first commissioning night, after alignment and system checks using stellar field plates were completed, a galaxy field plate was installed, and a 20 minute galaxy observation was made (Figure 4). Stacked images from subsequent nights observing were used to construct velocity fields for each of the target galaxies, see for example Figure 5.

## Conclusions

The first commissioning run of the SAMI instrument was extremely successful. This on-telescope experiment has demonstrated the potential of hexabundle technology for multi-object IFU science.

While some modifications to the instrument are required, it appears feasible to conduct dedicated science surveys with SAMI in the near future.

## Acknowledgements

We warmly thank all staff at the AAO for their support in developing and commissioning the SAMI instrument and the AAOmega spectrograph.

## References

Bryant, J. et al, MNRAS, 2011, arXiv1104.0985

# Galaxy Parameter Variations Across and Through the 6dFGS Fundamental Plane

Christina Magoulas (University of Melbourne), Christopher Springob (AAO), and the 6dFGS team

http://www.aao.gov.au/local/www/6df/

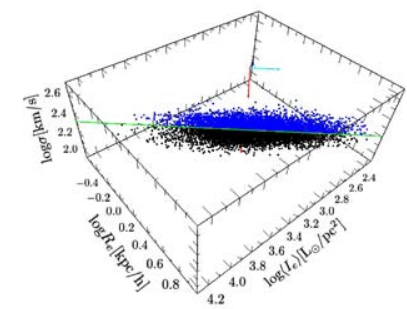

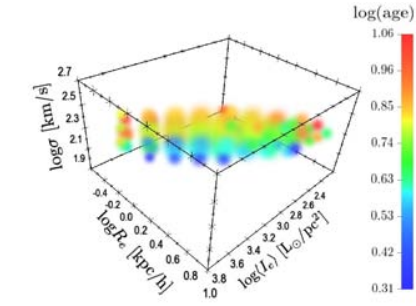

Figure 1: 6dFGSv galaxies plotted in (J-band) 'Fundamental Plane space'. Fundamental Plane space is the logarithmic parameter space determined by the three Fundamental Plane parameters, half-light radius $R_e$, velocity dispersion $\sigma$, and surface brightness $I_e$. The best-fit plane is shown in green.

Figure 2: 6dFGSv galaxies binned in FP space, with redder colours corresponding to increasing values of each stellar population parameter. The colour bar on the right shows the log of stellar age in Gigayears. The size of each sphere scales with the logarithm of the number of galaxies in the bin.

### References
Bernardi et al. 2003 AJ 125, 1866
Djorgovski & Davis 1987 ApJ 313, 59
Dressler et al. 1987 ApJ 313, 42
Graves, Faber, & Schiavon 2009a ApJ 693, 486
Graves, Faber, & Schiavon 2009b ApJ 698, 1590
Jarrett, Chester, Cutri, Schneider, Skrutskie, & Huchra 2000 AJ 119, 2498
Jones et al. 2009 MNRAS 399, 683
La Barbera, de Carvalho, de la Rosa, Lopes, Kohl-Moreira, & Capelato 2010 MNRAS 408, 1313
Magoulas et al., 2011, (in prep)
Pahre, de Carvalho, Djorgovski 1998 AJ 116, 1591
Saglia, Colless, Burstein, Davies, McMahan, & Wegner 2001 MNRAS 324, 389
Springob et al., 2011, MNRAS, (submitted)

From 2001 to 2006, the UK Schmidt Telescope undertook an all-southern sky spectroscopic survey, known as the Six-degree Field Galaxy Survey (6dFGS). The final data release of 6dFGS redshifts took place in 2009 (Jones et al., 2009), though we continue to make use of the survey's spectra for extragalactic and cosmological studies.

The peculiar velocity subsample of 6dFGS (6dFGSv) includes approximately 10,000 early-type galaxies. With the goal of ultimately deriving the distances and peculiar velocities of each of these galaxies, we make use of an important correlation for early-type galaxies, the Fundamental Plane (FP). The FP is a three-dimensional relation in the logarithmic space of galaxy size, velocity dispersion and surface brightness (Djorgovski & Davis 1987; Dressler et al., 1987). A redshift-independent distance to an early-type galaxy can be derived by measuring its offset from the FP relation. However, such an analysis requires us to carefully consider how we fit the FP, and how we account for selection effects. In this article, we describe how we fit the FP for this sample, and how we account for the trends of other parameters within 'FP space'. This process also allows us to gain new insights about galaxy formation and evolution.

### Sample Selection and Fundamental Plane Fitting

6dFGSv comprises 10,000 of the brightest galaxies with redshift z < 0.055 from the main redshift survey of 6dFGS. The near-infrared target selection of 6dFGS favours older, bulge-dominated galaxies, as these wavelengths are less sensitive to dust extinction than are optical wavelengths. The 6dFGSv sample combines 6dFGS velocity dispersion measurements with photometric parameters in the J,H and K near-infrared passbands from the Two Micron All Sky Survey (Jarrett et al., 2000).

Since the original formulation of the Fundamental Plane relation, the size and quality of galaxy samples has steadily increased (Bernardi et al., 2003; La Barbera et al., 2010; Graves et al., 2009a), yet the statistical models and fitting techniques used to determine the plane have not evolved with the same level of detail. Many previous studies used simplistic linear regression fitting techniques and model the distribution of galaxies in FP space as a 2D plane with scatter. However, these fitting methods fail to account for the censoring present in an observed FP sample and the measurement errors (and their correlations) in each of the FP parameters and so are susceptible to bias, limiting any interpretations drawn from their best-fit plane. Instead, we adopt a three-dimensional Gaussian model for the FP distribution, whose best-fit parameters are determined using a maximum likelihood method (in a similar manner to Saglia et al., 2001). Our fitting method properly accounts for selection effects in the sample (due to cuts in velocity dispersion and apparent magnitude) and observational error (and their correlations) in the FP parameters (Magoulas et al., in prep).

In the J-Band, our tightest sample, the best-fit Fundamental Plane has a slope and scatter that is consistent with other near-infrared Fundamental Planes (La Barbera et al., 2010; Pahre et al., 1998). The FP slope in velocity dispersion, a, is observed to be steeper (a ~ 1.5) in the near-infrared than is usually found in the optical (a ~ 1.2), which is consistent with an old stellar population. Figure 1 shows the distribution of galaxies in 6dF J-Band FP sample, in the space of the Fundamental Plane parameters along with the best-fitting plane of the three dimensional Gaussian.

### Morphology and the Fundamental Plane

We have visually inspected our entire sample of galaxies to classify their morphological type, identifying any disc-dominated galaxies that may be contaminating our sample. We have separated the sample into pure early-type galaxies (i.e., elliptical and lenticular) and spiral bulges. We have fit a Fundamental Plane to each of these morphologically selected subsamples individually. We find that they occupy the same FP, though there is some segregation *along* the FP, such that spiral bulges are preferentially larger in size. However, this trend is likely an artifact of our galaxy selection. We only include spiral galaxies whose bulge is enclosed by the large 6dF fibre. Therefore the spiral bulges in our sample are more likely to be galaxies that are larger and/or nearby. In any case, we find no evidence of a large offset of the FP between morphological subsamples. Including spiral bulges in our sample does not increase the total scatter of the Fundamental Plane.

### Stellar Population Parameters and the Fundamental Plane

The most recent FP studies (La Barbera et al., 2010, Graves et al., 2009b) have focussed on analysing trends of the FP with stellar population parameters in order to understand how they contribute to the scatter and tilt of the FP. We have derived stellar population parameters for 7132 of the galaxies in 6dFGSv, and examined the variation of the stellar population parameters in FP space. We bin the galaxies along the axes of the three-dimensional Gaussian model, fit with the maximum likelihood method discussed above, and compute median values of these bins of stellar age, [Fe/H], [Z/H], and [α/Fe]. We find clear trends in each of these parameters across and through the plane (Springob et al., 2011).

From the perspective of using the FP as a distance indicator, the most important of these trends is the trend of stellar age in FP space. As Figure 2 shows, stellar age is seen to vary directly through the plane. We find that selecting on age yields a modest reduction in the FP scatter, as galaxies younger than 3 Gyr occupy a plane with a scatter along the distance dimension of 31.5%, while those older than 3 Gyr occupy a plane with scatter 27.2%. In the coming months, we will be calculating peculiar velocities for all 10,000 galaxies in 6dFGSv, and accounting for these age trends to get more accurate distances.

# Dragonfly flutters its wings

Nemanja Jovanovic (AAO/Macquarie University), Barnaby Norris (University of Sydney), Simon Gross (Macquarie University), Paul Stewart (University of Sydney), Ned Charles (University of Sydney), Peter G. Tuthill (University of Sydney), Sylvestre Lacour (Observatoire de Paris), Martin Ams (Macquarie University), Jon Lawrence (AAO/Macquarie University), Graham D. Marshall (Macquaire University), Gordon Robertson (University of Sydney), Michael Ireland (AAO/Macquarie University), and Michael J. Withford (Macquarie University)

Since the first detection of exoplanets some 20 years ago (Wolszczan et al, 1992), more than 500 planets have been discovered to date (see http://exoplanet.eu). The majority of planets found have been detected via two methods: high precision surveys of stellar radial velocity (Wright et al., 2007), or transits in which the planet partially occults the light from its host (Borucki et al, 2010). Both techniques are strongly biased towards finding planets with higher masses in tight orbits. For studying planets that are further out there is strong interest in so-called "direct detection" techniques, such as coronography or optical stellar interferometry in which planetary light is isolated, negating the need for continuous monitoring over one or more orbital periods (>20 years).

With the aim of developing an instrument that allows for high dynamic range imaging for use in exoplanetary science, the Dragonfly instrument was conceived. Dragonfly is an optical stellar interferometer but unlike interferometers before it, it takes a revolutionary new approach to the design of such instrumentation; it is based on integrated photonic circuits (micro-optic chips), micro-mechanical mirrors and micro-optics as shown in Figure 1 (Tuthill et al, 2010).

The light enters the instrument from the left of Figure 1 and is directed onto a segmented mirror which is placed in the pupil plane of the telescope. This mirror can be used to steer each segment of the pupil in both tip and tilt and can be pistoned to alter the path length difference for each segment through the instrument. The light then passes through a beam reducing telescope which re-images the pupil onto a micro-lens array such that there is a 1-to-1 correspondence between each element on each component, respectively. The focused beams are then injected into an array of optical waveguides embedded within a photonic chip. The waveguides remap the pupil of the telescope into a 1-D slit at the output, which is known as pupil remapping. The emanating beams are recollimated with a second micro-lens array before being dispersed and focused onto an InGaAs detector.

This unique photonic based instrument offers many benefits including:

- **Spatial filtering** - the light propagating within the single-mode waveguides has a simple planar wavefront. As a result, all residual phase aberration across each sub-aperture is rejected, resulting in a mode-cleansed interferogram.

- **Simple, high precision optical path length matching** – Routing of waveguides within the chip can be achieved with sub-micron precision allowing for extremely precise matching of optical path lengths.

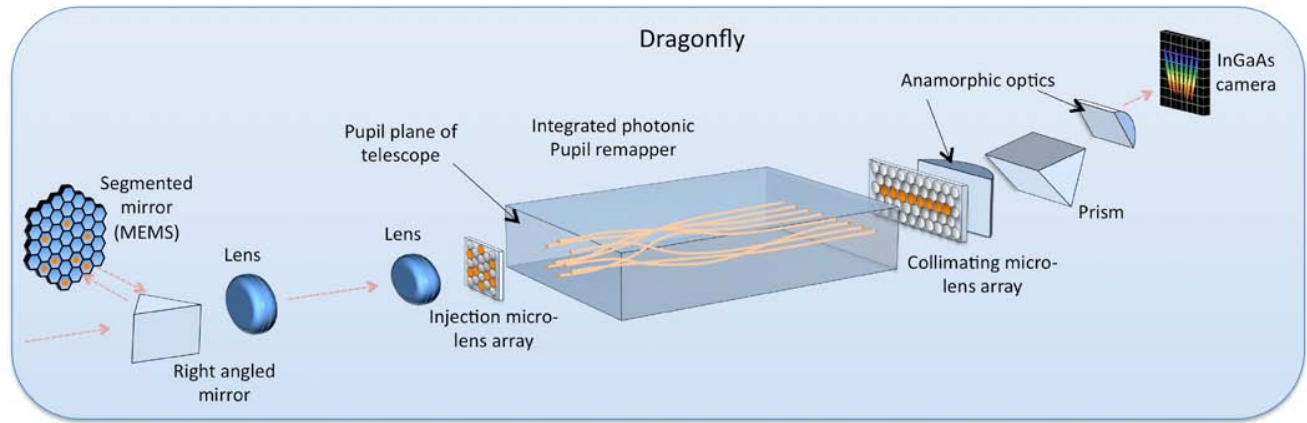

Figure 1: Schematic diagram of the Dragonfly instrument. Orange circles indicate the segments used on each optical component. Red arrows depict the direction of propagation of the stellar signal through the instrument, arriving from the telescope at the left.

- **Compact and robust** – An entire photonic chip encoding advanced processing capability will only measure several centimetres in size, and is thus a compact, robust device highly resilient to errors induced by thermal or mechanical anisotropies.

- **Non-redundant beam combination** – The input pupil geometry can be arbitrarily remapped into any output configuration. Therefore beam recombination schemes can be configured to eliminate any redundancy noise.

- **Full utilization of the available telescope pupil** – In contrast to aperture masking interferometry where very few holes are used in a plate to achieve non-redundancy, the remapping capability inherent to optical waveguides makes it possible to fill the entire pupil plane with waveguides, allowing a significant increase in sensitivity.

- **Cross-dispersed integral field unit** – Pupil remapping into a linear pattern permits the orthogonal ordinate of a 2-D array detector to be utilized to record spectral data.

- **Waveguide coupling optimization** – By positioning the segmented mirror and the injection micro-lens array at the pupil plane of the telescope, it is possible to use the tip/tilt feature of the segmented mirror to carefully optimise the coupling into the 10 μm optical waveguides.

The principle of the interferometer is that the output of each waveguide acts like a slit. By choosing which waveguides/slits have light injected into them, it is possible to choose the separation between slits and hence frequency of fringes that are generated on the detector as shown in Figure 2. It's not only possible to select a single set of fringes corresponding to a single baseline of the instrument as depicted in Figure 2(a) and (b), but it is also possible to turn on a combination of waveguides such that multiple sets of fringes with various fringe frequencies are superimposed on the detector simultaneously, as depicted in Figure 2(c). By measuring the phase associated with multiple fringe frequencies/baselines/spatial frequencies, it is possible to calculate a "closure phase", which is a measurable used to eventually reconstruct an image of the stellar target.

A prototype Dragonfly instrument was taken to the Anglo-Australian Telescope (AAT) and tested on the 20th and 21st of May, 2011. The entire instrument was installed in the UCLES pre-slit Coude room (Figure 3). The instrument was aligned to be co-linear with the injection beam to UCLES such that the UCLES slit viewing camera could be used for telescope pointing and tracking. For the on-telescope tests, an 8 waveguide pupil remapping photonic chip was used which reformatted the light into a plane of equally spaced waveguides. In order to eliminate the redundancy induced by having equally spaced waveguides, certain guides were switched off, in order to simulate a non-redundant array of guides as shown in Figure 2(c). This arrangement of waveguides was used as it allowed us to simulate 6 baselines simultaneously (i.e. 6 sets of fringes are overlapped on the detector at once).

Within 30 minutes after first injecting light from the scope, we had star light on the detector. This in itself is an accomplishment as the AAT does not have adaptive optics, making coupling into single-mode waveguides very difficult. Further, in order to avoid blurring of the fringes as a result of atmospheric fluctuations, we had to integrate at times commensurate with the atmospheric coherence time (~10 ms) which meant we had very low signals on a high read-noise laboratory grade detector.

On the second night, we pointed the telescope at Antares as it is one of the brightest objects in the sky in the H-band (~3.5mag) and after stacking frames for several hours we managed to demonstrate that we could detect all 6 baselines of the interferometer. Figure 4 shows the spatial frequency/baseline separation plot as a function of wavelength.

This was a fantastic proof-of-concept demonstration and over the next 6 months we will be refining Dragonfly ready to go to a larger scope with an adaptive optics system, where we hope to begin doing science with it. One day, Dragonfly may help astronomers hunt for and study faint exoplanets which will give us a better understanding of planetary and solar system evolution, but for now it's just a lot of fun. ✦AAO✦

### References

Borucki, W. J. et al, 2010, Science, 327, 977

Tuthill, P. et al, 2010, Proc SPIE, 7734, 51

Wolszczan, A. et al, 1992, Nature, 355, 145

Wright, J. T. et al, 2007, ApJ, 657, 533

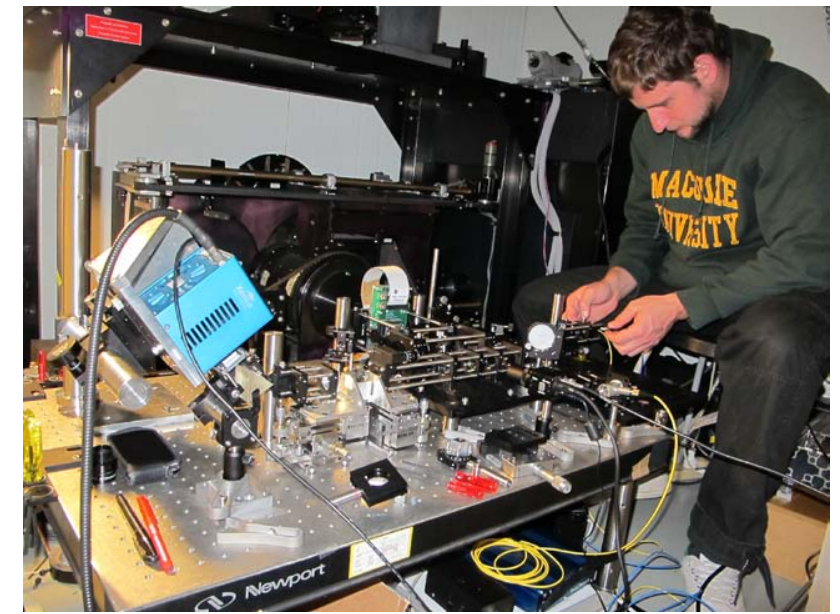

Figure 3: Nem aligning Dragonfly in the Coude room.

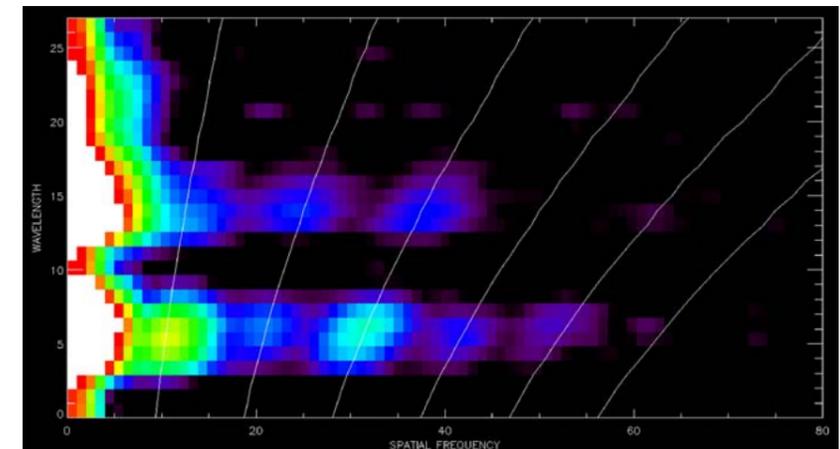

Figure 4: Power spectrum of the fringes detected by the system. Vertical axis is the wavelength decreasing from ~1.75 to 1 μm as you move up the axis. Horizontal axis shows the spatial frequency components of the interferogram. The lower band of dots is in the H-band and 6 can clearly be seen. The white lines show the predicted position of the spatial frequencies/baselines for the system.

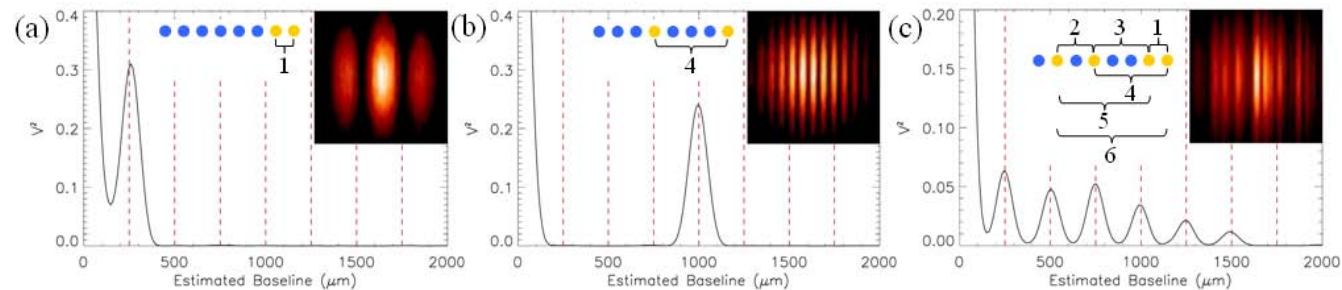

Figure 2: (a) and (b) depict the power spectra and fringe patterns (insets) associated with 2 pairs of waveguides with different separations. (c) Shows the power spectrum and the corresponding fringe pattern when light was injected into a combination of 4 waveguides. The dots represent the waveguides. Yellow – light injected, blue – light decoupled. The various baselines achieved with each combination of waveguides is displayed below each set of dots.

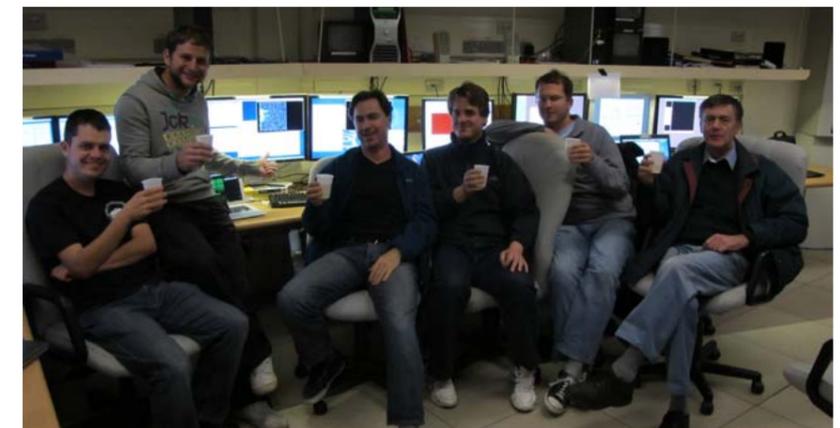

Figure 5: The Dragonfly team celebrating their fringe victory.

# RAVE at half a million

Fred Watson [AAO] and the RAVE collaboration

Back in June 2005, Coonabarabran laid on a sparkling winter weekend when the AAO hosted RAVE's seventh annual collaboration meeting. Now, the town has put on a repeat performance, with the thirteenth meeting being held in June 2011, again with AAO sponsorship. Between these two gatherings, RAVErs met in Ljubljana, Strasbourg, Baltimore, Padua and Groningen. Who says Coonabarabran can't mix it with the big league?

The 2005 meeting was held on the eve of RAVE taking over all available time on the UK Schmidt Telescope on a continuing user-pays basis. The project had been running in pilot mode for two years, and had already more than doubled the number of previously measured stellar radial velocities. By 2011, that number had been increased by a factor of ten, with 420,000 velocities in RAVE's internal database.

Perhaps more significant, though, was the total number of RAVE spectra obtained at the telescope, which, on the first day of the meeting, stood at 501,049. Reaching the half-million milestone endowed the whole event with an air of celebration that readily made the transition to the evening's conference dinner.

These impressive totals have not been arrived at without a huge effort from many people. Even before the observers face their nightly confrontation with 6dF and its ill-tempered robot, the target stars have to be selected and assembled into an input catalogue that is correctly formatted for the configuration software. Then there is the small army of AAO technicians who keep the elderly Schmidt Telescope and its fibre system operating at a level commensurate with RAVE receiving the best possible value for money.

From the telescope, the data are transferred to Macquarie University for quality control, and thence to Padua, where the calibration pipeline operates. The data next go to the Leibniz-Institut für Astrophysik Potsdam, where they are put through the radial velocity and atmospheric parameters pipelines, and eventually archived. A dataset like RAVE's demands extensive calibration and associated modelling, and this has involved scientists at several other institutions, including ANU, Johns Hopkins and the University of Ljubljana. Other centres involved in the project are l'Observatoire astrophysique de Strasbourg, the Open University, and the universities of Cambridge, Central Lancashire, Groningen, Heidelberg, Oxford, Sydney and Victoria.

RAVE spectra cover the calcium triplet region, and include stars with 8<*I*<12. Velocity precision is everywhere better than 2 km/s, but errors in $T_{eff}$, log $g$ and $Fe/H$ are dependent on magnitude. Three public data-releases have been issued to date, and are available on the RAVE website: www.rave-survey.aip.de/rave/. Three further releases are planned. DR4, containing data to 2008, will include distances and abundances; DR5 will include data to 2010, and the Final Data Release will contain all data obtained to the end of the project's observational phase, including spectra.

RAVE has produced more than 25 science papers since 2005. Over the past year, 12 refereed papers and three PhD theses have been accepted. They represent a wide range of galactic science, including high-velocity stars, multiple stars, stellar

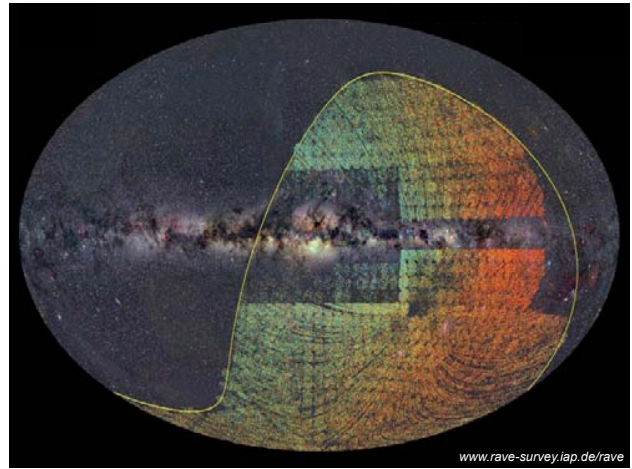

Galactocentric projection showing data from the current RAVE internal catalogue (approx. 420,000 stars), colour-coded for radial velocity in the range -50 km/s (blue) to +50 km/s (red). The reflex solar motion is hard to miss. [Credit: RAVE; background image (c) 2000 Axel Mellinger.]

distance determinations, thin and thick disc formation mechanisms, metal-poor stars, eccentricity distribution in thick disc stars, and new stellar streams.

At the 2011 meeting, science topics such as these jostled for space with progress reports and calibration procedures, and all presentations are now available for download by clicking on the participants' names at www.aao.gov.au/conf/RAVE/Participants.html. Also discussed were strategies for getting the best out of what is likely to be RAVE's final year of observations. The meeting ended with a visit to the telescope that has endowed the project with half a million spectra, together with an enchanting half-hour of southern hemisphere stargazing for its predominantly northern hemisphere participants.

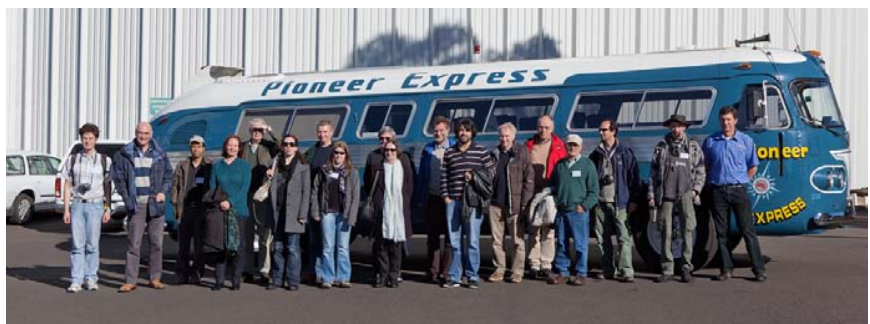

A band of intrepid RAVErs touring Siding Spring Observatory aboard the Acacia Motor Lodge's 1955-vintage coach. [Credit: Gal Matijevic]

# The Magellanic Quasars Survey

Szymon Kozlowski (Warsaw University Observatory),
Christopher S. Kochanek (Ohio State University)

The variability of quasars has provided many clues about the nature of active galactic nuclei (AGN), but recent developments are leading to a revolution in quantitatively analyzing variability. Variability time scales led to some of the first crude constraints on the sizes of AGN at all wavelengths. However, since few quasars were well monitored, studies focused on the statistical variability of large numbers of sparsely monitored quasars as a function of time scale. The largest analyses used tens of thousands of SDSS quasars, finding that quasar variability increases towards shorter optical wavelengths, lower luminosities and, probably, lower black hole masses (e.g., Van den Berk et al. 2004, de Vries et al. 2005). Detailed studies of individual quasars showed that they get bluer as they become brighter, consistent with thermal emission from a disk (e.g., Giveon et al. 1999). The most important application of these detailed studies, however, has been the development of "reverberation mapping" where the lag between continuum and line variations is used to measure the physical radius of the line emitting regions and then to calibrate relations between emission line widths and black hole masses (e.g., Peterson 1993).

But what about the light curves of individual quasars? How do you turn an apparently random wandering in brightness into numbers that you can analyze and use as astrophysical tools? A solution was proposed by Kelly et al. (2009) who showed that individual light curves could be well modeled as a particular stochastic process, the damped random walk (DRW). This reduces the random wanderings of the light curve into two numbers, an amplitude and a time scale, and Kelly et al. (2009) showed using a very small sample of about 100 AGN that these parameters appeared to be correlated with the physical properties of the AGN. In MacLeod et al. (2010), we expanded these studies to the largest available sample of quasars with light curves, the roughly 9000 quasars in the SDSS Stripe 82, finding that the time scales and amplitudes are correlated with rest-wavelength, luminosity and black hole mass.

Unfortunately, while the SDSS Stripe 82 quasar sample was the largest and best available, the light curves are not very good -- too few epochs and an effective duration too short to well-constrain the variability parameters of individual quasars. There are, however, quasars with extremely well-sampled light curves – any quasars lying in the fields monitored by the microlensing surveys of the

Magellanic Clouds and the Galactic bulge. The OGLE survey, for example, currently contains 15-year-long continuously growing light curves with ~1000 epochs for about 50 million sources in the Magellanic Clouds. For a serious study of variability, you need relatively large numbers of quasars since there are a minimum of three relevant physical variables: rest wavelength, luminosity and black hole mass (estimated from line widths). The necessary numbers of quasars are present in the microlensing fields with one little problem -- there are only about 25 quasars for every million LMC/SMC stars in these fields.

Standard optical color selection methods do not work for the Clouds, but there are two alternative methods. First, objects with red mid-IR colors (3.6 versus 4.5 microns) are either AGN or the relatively rarer stars with warm dust -- normal stars and galaxies are blue sources in these bands. The primary contaminants are AGB stars, young stellar objects (YSOs) and planetary nebulae (PNe), but these can be controlled with a few ancillary cuts, as we explored in Kozlowski & Kochanek (2009). Second, we can use the variability parameters for the SDSS quasars to identify sources varying like quasars. Here the contamination comes from irregularly varying massive stars, but most of the quasar parameter range differs from that of the stars, as we showed in Kozlowski et al. (2010). In both cases, you can define selection criteria that produce high purities at the price of reduced completeness.

When you go to plan your AAOmega observation, however, you realize you should simply forget about purity and aim for completeness because the density of available fibres is comparable to the density of all vaguely plausible quasar candidates. Tossing in all the reasonable mid-IR and variability-selected candidates and then adding possible OGLE matches to X-ray sources in the fields only comes to about 80 sources per square degree compared to an expectation of roughly 25 quasars per square degree with I<21 mag. Since there is no point in leaving a fibre empty, you should just do everything, including objects starting to be a little faint for the planned observation times. In this "no fibre left behind" approach, contamination is a feature, not a bug, and it will identify many peculiar LMC sources as a by-product.

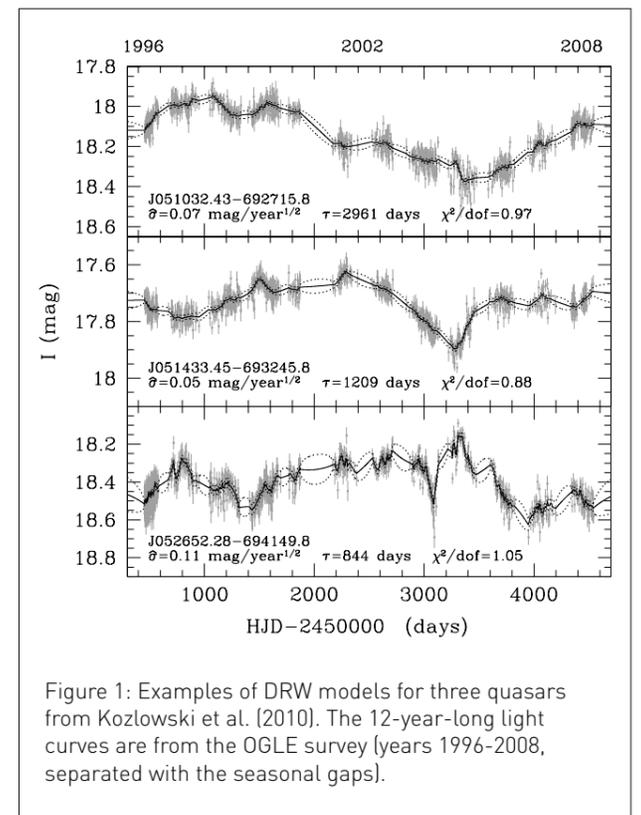

Figure 1: Examples of DRW models for three quasars from Kozlowski et al. (2010). The 12-year-long light curves are from the OGLE survey (years 1996-2008, separated with the seasonal gaps).

So far we have observed 4 AAOmega fields in the LMC and 1 in the SMC out of 17 planned, mostly while fighting cyclone Yasi and with only one field completed to the planned depth. In 5 hours on sky, we observed 1100 candidates, finding 174 new quasars, quadrupling (doubling) the number known behind the LMC (SMC). The quasar samples are roughly 80% complete for I<19.2 mag, and the completeness then drops. There are a comparable number of contaminating stellar sources, mostly previously unidentified B(e) stars, YSOs and PNe. The contamination and failure rates are high, but this was expected. We estimate that completing the current OGLE-III LMC and SMC fields would yield a total of ~700 quasars, and that an additional 3600 quasars can be identified in the OGLE-IV fields, ignoring any potential gains from obtaining the longer, planned integrations. While the samples would still be smaller than the ~9000 SDSS quasars we considered in MacLeod et al. (2010), the longer and more densely sampled OGLE light curves should let us examine the physical correlations of the variability parameters with a higher accuracy, making it easier to determine the intrinsic parameter distributions and their correlations with the physical properties of the quasars.

These quasars have additional uses beyond studies of quasar variability physics. They supply a dense network of reference points for improving the proper motion measurements, particularly since the motions of the Clouds are presently limited by the need to map out the internal motions of the Clouds. The brighter quasars can also be used for absorption line studies of the interstellar medium, where we have already identified 10 new quasars brighter than I<18 mag.

**References:**
de Vries et al. 2005, AJ, 129, 615
Giveon et al. 1999, MNRAS, 306, 637
Kozlowski & Kochanek 2009, ApJ, 701, 508
Kozlowski et al. 2010, ApJ, 708, 927
Kozlowski et al. 2011a, ApJS, 194, 22
Kozlowski et al. 2010b, arXiv:1106.3110
MacLeod et al. 2010, ApJ, 721, 1014
Peterson 1993, PASP, 105, 247
Van den Berk et al. 2004, ApJ, 601, 692

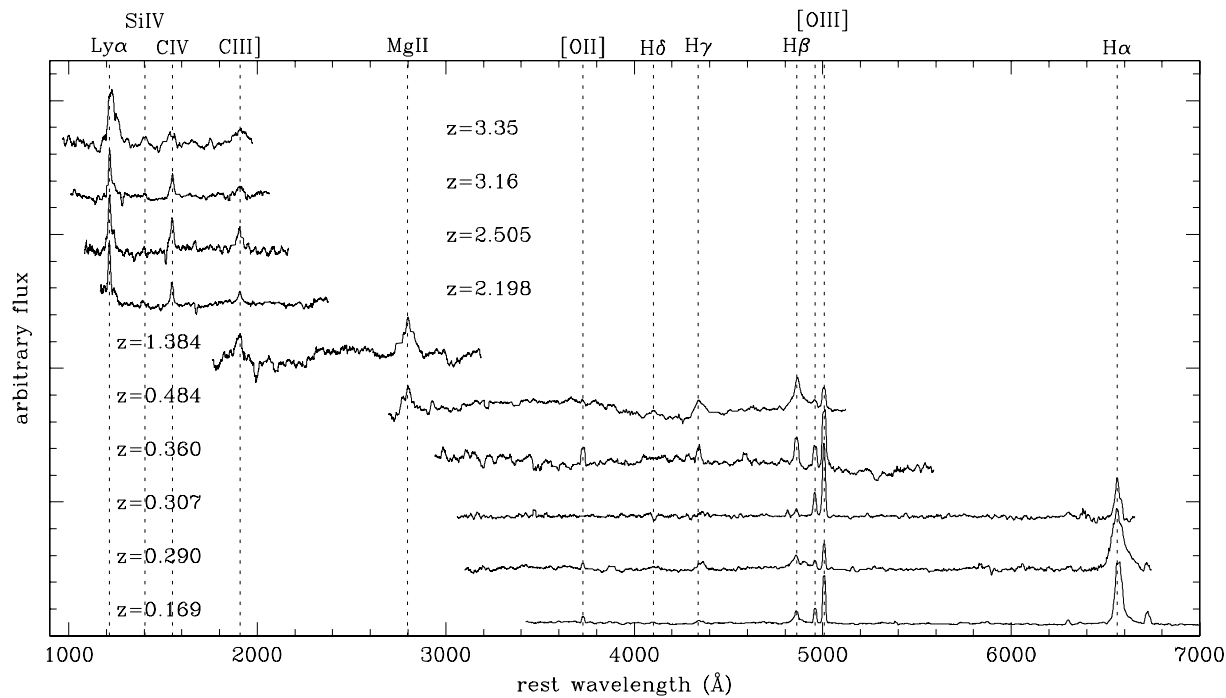

Figure 2: The spectra of ten of the 145 new LMC AGNs from Kozlowski et al. (2010b). The spectra have been flattened, smoothed and scaled, and emission lines from the ISM of the LMC have been suppressed.

# The Integrated Photonic Spectrograph's First Look at the Heart of the Scorpion

Nick Cvetojevic (Macquarie University), Nemanja Jovanovic (Macquarie University/AAO), Joss Bland-Hawthorn (University of Sydney), Roger Haynes (Astrophysics Institute Potsdam), Mick Withford (Macquarie University), and Jon Lawrence (AAO/Macquarie University)

At around 2am on a cold, clear night in mid May, light from the star Antares ended its six hundred year journey by bouncing off the AAT's mirror and into an integrated photonic spectrograph, the first of its kind in the world. The first successful detection of a stellar spectrum by our photonic spectrograph, accompanied by the popping of champagne bottles, was a major achievement for astrophotonic technology as well as bolstering the AAO's prominence in new and innovative instrumentation for astronomy.

The power of astrophotonics to expand the capabilities of astronomical instruments by exploiting photonic technologies has been recently illustrated by a number of demonstration devices (see Bland-Hawthorn & Kern 2009). The Integrated Photonic Spectrograph (IPS), essentially a complete spectrograph on a single photonic chip, is one such device (Figure 1a). While several alternative technologies for miniature astronomical spectrographs have been proposed (e.g., Watson 1996, Bland-Hawthorn & Horton 2006), we have concentrated on the planar arrayed-waveguide grating (AWG). In an AWG spectrometer (Smit & Van Dam 1996), light is coupled from an input fibre, through a free-propagation zone of constant refractive index, into a series of waveguides with a constantly increasing length increment. The output of these grating waveguides interferes within a second free-propagation zone producing wavelength dispersion, and is focused on the output face as a spectrum (Figure 1b).

Four key factors outline the potential advantages of an integrated photonic approach over the traditional spectrograph design, in particular for multi-object spectroscopy:

- **miniaturisation:** the entire spectrograph unit, including collimating and focusing optics and the disperser, is contained in a single, silica chip a few centimetres wide and a few millimetres thick. With such a significant reduction in size, very large multiplexing capacity (e.g., more than a hundred thousand spatial elements) becomes feasible;

- **integration:** with all the components of a classic spectrograph integrated on a single chip, we diminish the impact of problems such as alignment, stability and flexure which often add complexity to instrument design. Moreover, it permits exciting new possibilities for space-based or high temperature-fluctuating applications.

- **mass production:** by relying on mature lithographic fabrication techniques developed principally by the telecommunication industry we can ensure that instrument costs are not completely overwhelmed by technology development and demonstration costs. Furthermore, as the primary expense in AWG chip fabrication is the up-front tooling cost, it becomes practical (if not preferable) to fabricate large quantities;

- **modularisation:** by breaking the system into low-cost modular components, low-maintenance, readily-expandable instruments can be constructed with a large number of elements.

The IPS was initially demonstrated on-sky at the AAT with an observation of the emission spectrum of OH from the atmosphere (Cvetojevic et al. 2009). For that initial test, the IPS device was not interfaced to the telescope but was simply collecting the light directly from the sky. The next logical step was to build an instrument that can successfully take the light from the AAT and observe a spectrum from an astronomical source.

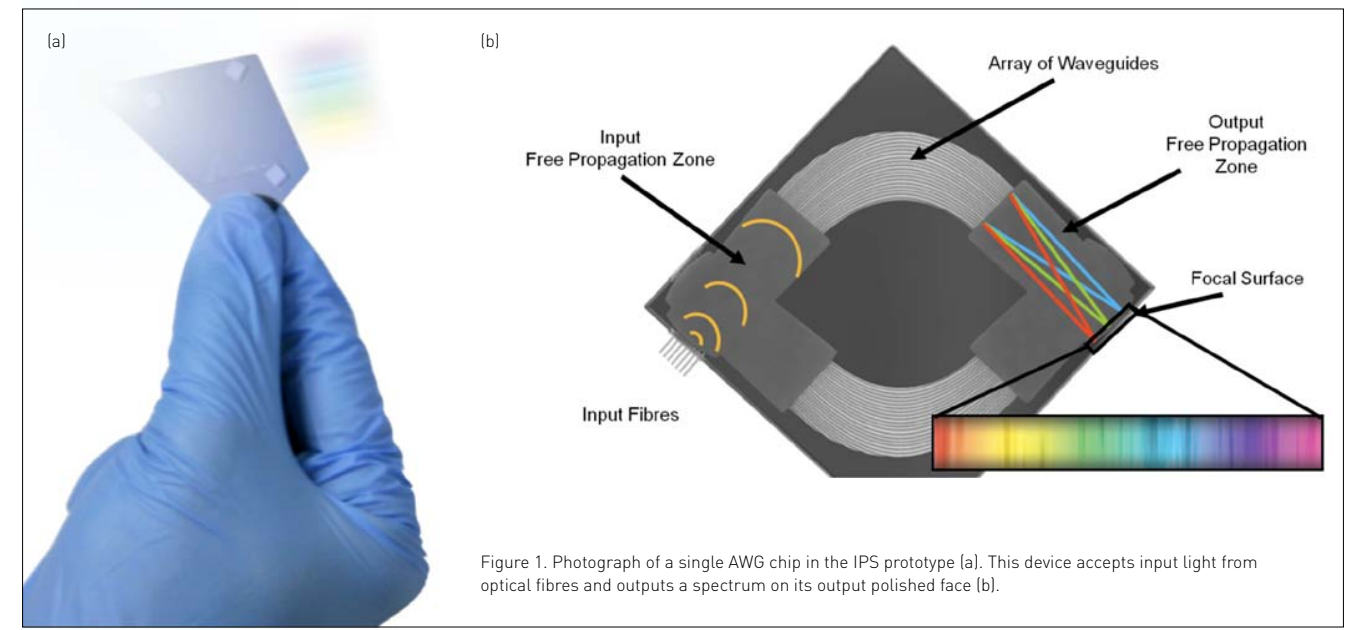

Figure 1. Photograph of a single AWG chip in the IPS prototype (a). This device accepts input light from optical fibres and outputs a spectrum on its output polished face (b).

# Using AAOmega to measure the age of the young open cluster IC2602

Paul Dobbie (AAO), Nicolas Lodieu (IAC) and Rob Sharp (ANU)

On the 19th of May, we interfaced a new IPS prototype with the AAT using an IFU which fed a multimode fibre (a fibre with a 50 micron core) and in turn a photonic lantern. This is the same interface as used by the GNOSIS OH-suppression instrument. The photonic lantern acts as a multimode-to-single mode converter, allowing our diffraction-limited IPS (inherently single-moded) to collect the light propagating down the fibres. Furthermore, we modified the IPS to simultaneously input twelve single-mode fibres, thus increasing our observational efficiency by an order of magnitude. The IPS chips have a throughput of ~80% and a typical resolving power of R=7000. The output face of the chips was imaged onto the IRIS2 detector, which we used in spectroscopic mode as a cross disperser for the astronomic H-Band (1480-1825nm), in which we were observing.

The first spectrum of the night was of Antares (α Scorpii) as it was the brightest sky up at the time, and served as a calibration source for final alignment adjustments (Figure 2). With a world-first under our belts, we continued to observe a Be star (alf Ara) (Figure 3a) and V* Pi 01 Gru, a cold red giant (Figure 3b), as they would hopefully contain more interesting spectral features in the H-band.

This demonstration is, to our knowledge, the first time a photonic spectrograph has successfully taken spectra of a source beyond the Earth, and the first time such a device was interfaced to a telescope. With the success of the new prototype, we have shown the practicality of such a device for astronomy. Further work on the IPS continues with potential interfacing with adaptive-optics aided telescopes, and a large redesign and fabrication of new AWG chips more ideally suited for astronomy. While still a long way from becoming a fully-fledged instrument, the recent successes make our ambition of having a spectrograph for astronomy that you can fit in your hand, seem very bright indeed.

### References

F.G. Watson, (ed. S.C.Barden), Proc.S.P.I.E., 2476, 68–74, 1995.

Bland-Hawthorn, J., & Horton, A., Proc. SPIE, 6269, 62690N, 2006

Bland-Hawthorn, J., & Kern, P., Opt. Exp., 17, 1880, 2009

Cvetojevic, N., Lawrence, J.S., Ellis, S.C., Bland-Hawthorn, J., Haynes, R., & Horton, A., Opt. Exp., 17, 18643, 2009

Smit, M.K., & Van Dam, C., IEEE J. Sel. Topics Quantum Electron., 2, 236, 1996

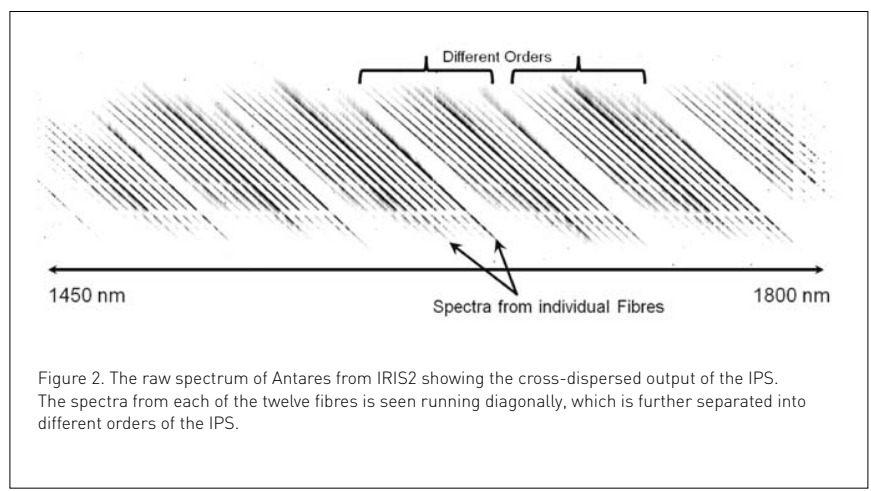

Figure 2. The raw spectrum of Antares from IRIS2 showing the cross-dispersed output of the IPS. The spectra from each of the twelve fibres is seen running diagonally, which is further separated into different orders of the IPS.

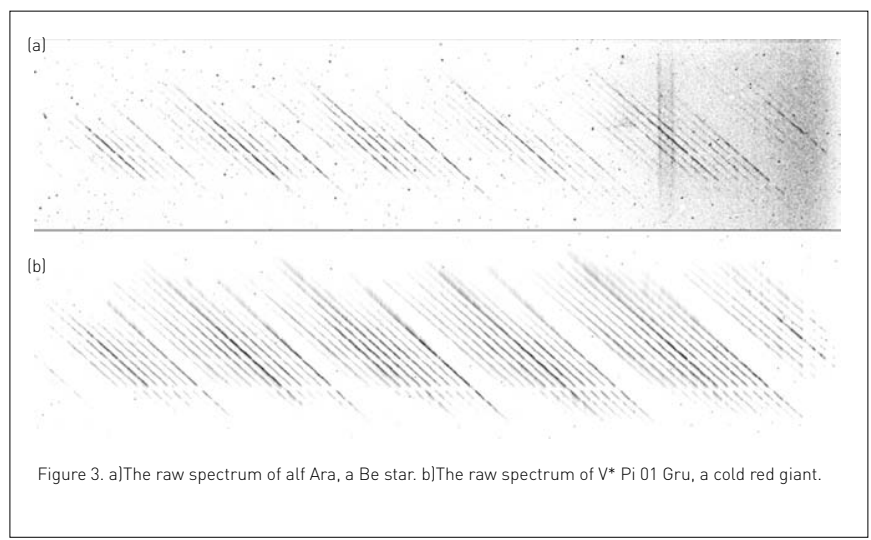

Figure 3. a)The raw spectrum of alf Ara, a Be star. b)The raw spectrum of V* Pi 01 Gru, a cold red giant.

Open clusters consist of co-eval populations of stars residing at similar heliocentric distances, that have formed from molecular gas clouds with near uniform chemical compositions. The common properties of their members render them excellent targets for addressing fundamental questions in stellar astrophysics. For example, they are frequently utilised to investigate the forms of the stellar/substellar initial mass function (e.g. Moraux et al. 2005; Lodieu et al. 2009) and the stellar initial mass-final mass relation (e.g. Williams et al. 2004; Kalirai et al. 2005; Dobbie et al. 2009). Open cluster members are also frequently used to probe stellar magnetism (e.g. Marsden et al. 2005), the evolution of stellar angular momentum (e.g. Irwin et al. 2009) and the mixing processes which occur within stars (e.g. Pinsonneault 1997).

Critical to the success of many of these types of investigation is the availability of a reliable age determination for the population under scrutiny. This allows meaningful comparisons to be drawn between the measured properties of the members of different clusters and aids in the judicious interpretation of the observed trends in the theoretical context (e.g. the variation of Li abundance in stars of different effective temperatures as a function of age; Randich et al. 2001). In general, age estimates for open clusters are obtained using the main sequence turn-off technique (MSTO), where the observed location in the luminosity-temperature plane of stars at the end of their main sequence life is compared to the predictions of stellar evolutionary models (Meynet, Mermilliod & Maeder 1993). However, for all but the oldest populations these age estimates are muddied by uncertainties associated with the extra convective mixing which is believed to occur at the boundary between the cores of stars and the overlying layers (Woo & Demarque 2001). The degree of convective core overshooting adopted in stellar evolutionary models impacts both the predicted main sequence lifetimes and the luminosities of stars, with greater overshoot leading to larger cluster age estimates. Thus a method of determining ages which is independent of assumptions about the physics of the stellar core boundary is to be preferred.

We are fortunate that such a technique has been successfully demonstrated (Basri et al. 1996). It relies on locating the boundary, in terms of mass, at which the element Li re-appears in the spectral energy distributions of the completely convective very-low-mass stellar and substellar members of a population (Rebolo et al. 1992). As a population matures, up to 300Myr, the location of this Li depletion boundary (LDB) migrates to lower mass (corresponding to later spectral types), providing a potentially excellent handle on the age (D'Antona & Mazzitelli 1994). However, the draw back of this approach is that it is reliant on the availability of moderate signal-to-noise, medium resolution optical spectroscopy of intrinsically faint red objects. Hence, prior to this work there were only five young open clusters with LDB based age determinations.

With the commencement of the operation of near-IR survey instruments such as VISTA, which are capable of probing the sequences of young, nearby stellar populations down to planetary masses, it is desirable to expand the set of southern open clusters for which LDB ages are available so that these data sets can be fully exploited. The nearby (d=150-170pc e.g. Braes 1961, van Leeuwen 2009) pre-main-sequence population, IC2602 is potentially one of the most astrophysically interesting of these targets. Currently, it's age, which is estimated to lie in the range t~30–70 Myr (e.g. Kharchenko et al. 2005), is rather poorly determined. As this cluster is located close to the Galactic plane (b=−4.9°) and spread over several square degrees of sky, it is rather difficult to distinguish members from reddened background field stars via photometric surveys. Therefore, assembling a reasonably clean list of faint candidate members for the (traditional) long-slit spectroscopic follow-up program, which is necessary to determine the location of the LDB, is extremely challenging. Indeed, despite the youth and proximity of IC2602, no low mass stellar or substellar members have been reported prior to this study. Fortunately, the difficulties above can be mitigated to a large extent by undertaking the follow-up observations with a high gain, multi-object spectrograph which has a wide field-of-view i.e. AAOmega. Here we briefly describe our recent work on IC2602 aimed at securing a LDB age for this population.

### The initial identification of candidate low mass cluster members

We retrieved I band imaging for ~1.4 square degrees towards IC2602 from the European Southern Observatory's (ESO) data archive. This had been obtained with the 2.2m telescope and the Wide Field Imager (WFI, Baade et al. 1999) during February 2001. These datasets were processed and calibrated using the Cambridge Astronomical Survey Unit's CCD reduction toolkit (Irwin & Lewis 2001). The optical detections were cross correlated with the Two Micron All-Sky Survey (2MASS, Skrutskie et al. 2006) point source catalogue to obtain corresponding J and K band magnitudes for the brighter sources. Subsequently, these data were used to construct an I-J,J colour-magnitude for objects classified as stellar, towards the cluster. As our guide to the likely location of the cluster sequence we used a selection of low mass stellar members of the Pleiades for which I and J photometry are available in the literature (Stauffer et al. 1994, 1998). The photometry of these Pleiades objects was shifted (from an assumed d=135pc; Pan et al. 2004) to the new Hipparcos distance of IC2602 and adjusted by 0.45 magnitudes to larger luminosities (based on the predictions of evolutionary models for low mass stars) to account for the expected lower age of this cluster (i.e. ~50Myr). Based on the minimum and maximum nuclear age estimates for IC2602 in the literature, we estimated the probable location of the LDB to be J~13.2-15.5 mag (corresponding to mid-M spectral types). We initially selected all sources within these magnitude limits and on or to the red of the line defined loosely by the Pleiades objects as candidate members of the cluster. After visually inspecting these sources to weed out blended stars, we were left with 249 for spectroscopic follow-up.

## Multi-object spectroscopy of candidate low mass members of IC2602

We used the AAOmega + 2dF facility (Sharp et al. 2006, Lewis et al. 2002) at the Anglo-Australian Telescope for a mere 2 hours during February 2010 to obtain medium resolution spectroscopy of 219 of our 249 candidate members of IC2602. These observations were performed using the 1000R grating on the red arm to cover the wavelength range 6200-7300 angstroms, within which the resonance line of Li is located. Our science observations were bracketed by fiber-flat and arc calibration frames. Simultaneous observations were obtained with the 580V grating on the blue arm but these are not relevant to this study. The red arm data were processed using a slightly modified procedure to the standard reduction implemented by 2DFDR. Due to the presence of weak but slowly variable H$\alpha$ emission across our AAOmega field-of-view we used a median of the background counts from the closest ~8 sky fibres to each science fiber to obtain an estimate of the night sky spectrum. An analysis of the residuals in each of the sky fibres showed a noticeable improvement to the background subtraction accuracy using this approach.

Out of the total of 219 objects observed spectroscopically we found only 22 had energy distributions that resembled mid-M spectral types. As young very-low-mass stars typically display strong [EW<–few angstroms, Barrado y Navascues et al. 2004] H$\alpha$ emission, based on our by-eye inspection of the AAOmega datasets, we conditionally labelled as candidate spectroscopic members 17 of these 22 sources. The vast majority of the spectroscopically observed objects that we rejected appeared to be of K or very-early M spectral types and were presumed to be reddened background stars, which are known to be a major issue for photometric studies of this cluster (e.g. Foster et al. 1997). To assess more quantitatively the spectral types of the 17 remaining candidates we computed spectral indices based on the strengths of the TiO and CaH bands (TiO5, CaH1, CaH2, and CaH3; Reid et al. 1995; Cruz & Reid 2002) which are present in the wavelength range covered by our AAOmega observations. As an additional check we also compared all the spectra to each other in order to define a spectral sequence and then matched them to low resolution spectroscopic templates of the young M3–M6 dwarfs members of the Chamaeleontis region (Luhman 2004; Luhman & Steeghs 2004). We found our objects to lie within the range M3.0 to M5.5. While a substantial proportion of field stars of these spectral types display H$\alpha$ emission, scrutiny of the large spectroscopic sample studied by West et al. (2008) suggests that only 15-20% of M4/M5 dwarfs within 50pc of the Galactic plane should be expected to have H$\alpha$ equivalent widths <–6°A. Since our spectroscopic investigation unearthed a total of 16 M4/M5 dwarfs, we found that relative to the field population our remaining sample contained a significant excess (10) of these strong H$\alpha$ emitting objects. Many of these sources were found to lie on a relatively tight sequence in the I-J, J colour-magnitude diagram, bracketed by 20Myr and 50Myr old theoretical isochrones (Figure 1).

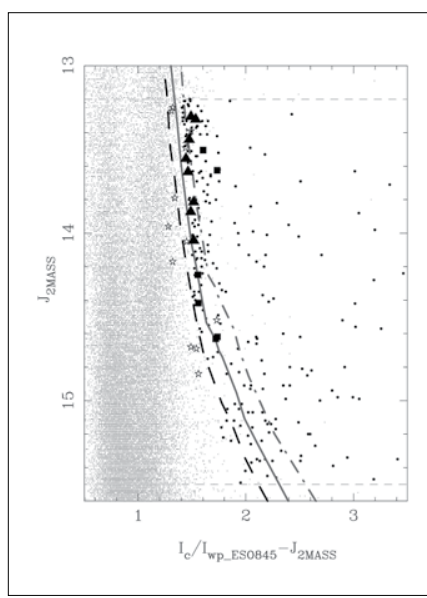

Figure 1: The I–J,J colour-magnitude plot of all point sources in the ESO WFI field-of-view which have a 2MASS counterpart. As a guide to the location of the cluster sequence, a selection of Pleiades low mass stars (open stars) and a NextGen theoretical isochrone for 50Myr (dashed line), both shifted to the distance of IC2602 have been overplotted. Objects that were selected for spectroscopy are highlighted (large dots) as are spectroscopic members with (filled squares) and without (filled triangles) lithium in absorption at 6707.8 angstroms 20Myr and 32Myr theoretical NextGen isochrones, adjusted as per the 50Myr model, are also shown (solid and dot-dashed grey lines respectively).

We also measured the radial velocities of our 17 candidate low mass candidates using the SPLOT routine in IRAF to fit a Voigt profile to the observed H$\alpha$ lines. The members of a star cluster display only a small dispersion (delta $v_r$<1kms$^{-1}$) about the cluster mean value. Indeed, Marsden et al. (2009) recently determined $v_r$=17.4±1.0kms$^{-1}$ from a high-resolution study of solar type cluster members. We estimated the uncertainties in our measurements to be ~7kms$^{-1}$ based on the dispersion seen in the radial velocities determined from each of the four unstacked spectra of a subsample of the candidates. We verified their internal accuracy by using the IRAF FXCOR routine to estimate the radial velocities of each targets with respect to 2MASS J10430236–6402132, the brightest object in our sample. We found a broad peak in the distribution of radial velocity measurements centered around 15kms$^{-1}$. This likely corresponds to the cluster population. We chose to select as radial velocity members those objects which were within 2$\sigma$ of the recent Marsden et al. (2009) measurement i.e. $v_r$=3.4-31.4kms$^{-1}$. This led to the rejection of a further three sources that as it turned out, had by at least 1.5 subclasses, the earliest spectral types (all are M3.0). Up to this point our combined selection criteria should have rejected almost 80-90% of mid-M field stars. If the remaining sample of 14 candidates was in fact dominated by field dwarfs (as opposed to cluster members) then, in our entire spectroscopic sample we should have unearthed ~80 objects of mid-M spectral type.

## Astrometry using 2dF's Focal Plane Imager

As one final probe of cluster membership status, we determined the relative proper motions of a number of candidates using the ESO WFI data and frames acquired with the Focal Plane Imager (FPI) of the AAOmega spectrograph. The FPI is an Apogee imaging camera which is mounted in front of the skyward facing field plate of 2dF. It contains a 512×512 pixel Kodak KAF-0261E CCD and has a field of view of 2.6´×2.6´. We acquired 3×10s exposures through a RG630 filter for 2MASS J10422712–6421401, 2MASS J10420463–6434373, 2MASS J10430890–6356228, 2MASS J10430236–6402132 and 2MASS J10432126–6419594 in the twilight prior to the regular AAOmega observing nights during May 2010. On these evenings, the sky was clear and seeing was substantially better than the Siding Spring median value (i.e. 0.9-1.2"). We reduced the images following standard procedures within the IRAF software environment, namely bias and dark subtraction, flat fielding, astrometric calibration and co-addition.

We used the IRAF routine DAOFIND to determine the positions of reference objects of comparable or greater brightness to our candidates in the two images. We cross-matched these lists of positions using the STARLINK TOPCAT software. Subsequently, we employed routines in the STARLINK SLALIB library to construct a six co-efficient linear transform between the two images of each candidate, where >3$\Sigma$ outliers were iteratively clipped from the fits. The proper motion, in pixels, was determined by taking the difference between the observed and predicted location of a candidate in the 2nd epoch dataset. This was then converted into milli-arcseconds per year in right ascension and declination using the world co-ordinate system of the 1st epoch imaging and dividing by the time baseline between the two observations (9.23 years). The relative proper motion vector point diagram for these objects (triangles with error bars) is shown in Figure 2. 2MASS J10422712–6421401, 2MASS J10420463–6434373, 2MASS J10430890–6356228 and 2MASS J10430236–6402132 appear clumped together very close to the location of known bright cluster stars, supporting our conclusion that these are members of IC2602. An additional source, 2MASS J10432126–6419594, lies well away from the four other objects but we had earlier discounted it on the basis of spectral type and radial velocity.

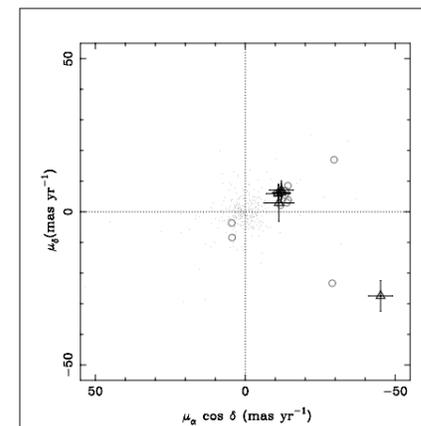

Figure 2: The vector point plot of the relative proper motions of several candidate members of IC2602 (black triangles with error bars). Overplotted are the UCAC3 proper motions of known bright cluster members (grey open circles). The relative proper motions of other objects within the FPI fields are also shown (small dots).

## The LDB and the age of IC2602

We examined the spectra of the 14 remaining candidate members of IC2602 for the presence of the 6707.8 angstrom Li absorption line. Compelling evidence of this feature was found in the spectra of six objects. We used the SPLOT routine in IRAF to measure the equivalent width of the line in these six spectra and also to set upper limits on the equivalent width in the remaining datasets where there is no convincing detection. Notably, we found two objects with clear detections of the Li line that are substantially brighter than the four other Li rich candidates. The location of these objects approximately +0.75 mag above the sequence defined by the other spectroscopic members could indicate that these two objects are near equal mass binary stars. We could have expected to detect one or two binary members in a sample of this size since the binary fraction of low mass field and open cluster stars is estimated to be 30% (Leinert et al. 1997; Pinfield et al. 2003). Of course, with the available data were unable to exclude the possibility that these are simply young field star interlopers but their spectral classifications are consistent with the binary hypothesis. Li was positively detected in the spectra of all four candidate members fainter than 2MASS J10430890–6356228 which has J=14.04±0.03 mag (K$_S$=13.17±0.04 mag). The brightest Li rich candidate with a position in the colour-magnitude diagram which was found to be consistent with the single star sequence defined by the other members is 2MASS J10439236–6402132 which has J=14.25±0.05 mag (K$_S$=13.30±0.04 mag). Following Manzi et al. (2008), on the basis of the photometry of these two objects we concluded that the LDB of IC2602 lies within the range J=14.01–14.30 mag. (K$_S$=13.13–13.34 mag), allowing for the uncertainties in the 2MASS photometry. Spectroscopy around the 6707.8 Å line for the four objects closest to either side of the LDB is shown in Figure 3.

We used the derived M$_J$ and M$_K$ magnitudes of the LDB and the theoretical near-IR photometry of Baraffe et al. (1998) to estimate the age for IC 2602. To derive these absolute magnitudes we adopted a distance modulus of (m-M)$_0$=5.86±0.1, which is based on the most recent Hipparcos determination and is consistent with the majority of other distance estimates in the literature (Whiteoak 1961; Braes 1961; Robichon et al. 1999; van Leeuwen 2009). We assumed reddening of A$_J$=0.03 and A$_K$=0.01 based on E(B-V)=0.035 (Hill & Perry 1969) and A$_J$/E(B-V)=0.86 and A$_K$/E(B-V)=0.36 (Fitzpatrick 1999). We also considered the finite depth of the cluster since it is effectively a further uncertainty on the distance to each individual member. We estimated that most low mass stars should lie within 4 pc of the cluster centre which corresponds to ±0.06 on the distance modulus of individual members. Adding the uncertainties in quadrature we found that the LDB in IC2602 occurs within the ranges M$_J$=8.03–8.49 mag and M$_K$=7.17–7.56 mag. These were compared to the theoretical predictions of the Lyon group (Chabrier & Baraffe 1997; Baraffe et al. 1998) for M$_J$ and M$_K$ as a function of age, at which 99% of a stars primordial Li has been burned (Figure 4). Based on these photometric bands we determined the age of IC2602 to be 46(+6/-5) Myr. This LDB derived age is larger by a factor 1.8 than the age (25Myr; Stauffer et al. 1997) obtained by comparing the position in the HR diagram of the low mass cluster sequence to the predictions of the models of D'Antona & Mazzitelli (1994). It is also greater by a factor 1.3 than the MSTO age of 35Myr determined by Mermilliod (1981) using stellar evolutionary models which allow for modest levels of convective core overshoot (Maeder & Mermilliod 1981). In contrast, Kharchenko et al. (2005) determined a MSTO age of 67Myr using the more recent evolutionary models of Girardi et al. (2000). However, their estimate is based on the location in the HR diagram of just two stars and should also be taken in the context of their larger age determination (75Myr) for the similar cluster, IC2391 (also based on just two stars).

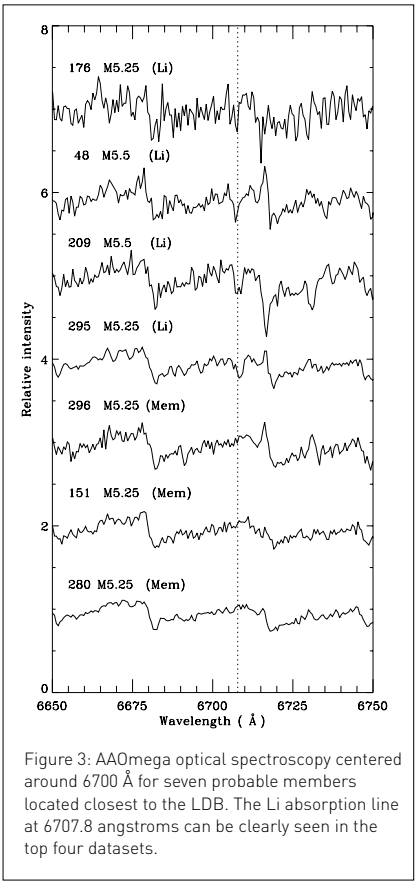

Figure 3: AAOmega optical spectroscopy centered around 6700 Å for seven probable members located closest to the LDB. The Li absorption line at 6707.8 angstroms can be clearly seen in the top four datasets.

We found IC2602 to be consistent with the general trend delineated by the Pleiades, α-Per, IC2391 and NGC2457, whereby the LDB age is 120-160% of the estimates derived using more traditional techniques. This trend is only currently bucked by IC4665 where the MSTO age determination (36Myr; Mermilliod 1981) is larger than the LDB age by a factor 1.3 (see Manzi et al. 2008).

### References

Baade D., Meisenheimer K., Iwert O., Alonso J., Augusteijn T., Beletic J., Bellemann H., Benesch W., 30 co-authors 1999, The Messenger, 95, 15

Baraffe I., Chabrier G., Allard F., Hauschildt P. H., 1998, A&A, 337, 403

Barrado y Navascués D., Stauffer J. R., Jayawardhana R., 2004, ApJ, 614, 386

Basri G., Marcy G. W., Graham J. R., 1996, ApJ, 458, 600

Braes L. L. E., 1961, Monthly Notes of the Astronomical Society of South Africa, 20, 7

Chabrier G., 2003, PASP, 115, 763

Chabrier G., Baraffe I., 1997, A&A, 327, 1039

Cruz K. L., Reid I. N., 2002, AJ, 123, 2828

D'Antona F., Mazzitelli I., 1994, ApJS, 90, 467

Dobbie P. D., Napiwotzki R., Burleigh M. R., Williams K. A., Sharp R., Barstow M. A., Casewell S. L., Hubeny I., 2009, MNRAS, 395, 2248

Fitzpatrick E. L., 1999, PASP, 111, 63

Foster D. C., Byrne P. B., Hawley S. L., Rolleston W. R. J., 1997, A&AS, 126, 81

13. Girardi L., Bressan A., Bertelli G., Chiosi C., 2000, A&AS, 141, 371

Hill C., Perry C. L., 1969, AJ, 74, 1011

Irwin J., Aigrain S., Bouvier J., Hebb L., Hodgkin S., Irwin M., Moraux E., 2009, MNRAS, 392, 1456

Irwin J., Hodgkin S., Aigrain S., Bouvier J., Hebb L., Irwin M., Moraux E., 2008, MNRAS, 384, 675

Kalirai J. S., Richer H. B., Reitzel D., Hansen B. M. S., Rich R.M., Fahlman G. G., Gibson B. K., von Hippel T., 2005, ApJL, 618, L123

Kharchenko N. V., Piskunov A. E., R¨oser S., Schilbach E., Scholz R.-D., 2005, A&A, 438, 1163

Leinert C., Henry T., Glindemann A., McCarthy Jr. D. W., 1997, A&A, 325, 159

Lewis I. J., Cannon R. D., Taylor K., Glazebrook K., Bailey J. A., Baldry I. K., Barton J. R., 16 co-authors 2002, MNRAS, 333, 279

Lodieu N., Zapatero Osorio M. R., Rebolo R.,Martín E. L., Hambly N. C., 2009, A&A, 505, 1115

Luhman K. L., 2004, ApJ, 616, 1033

Luhman K. L., Steeghs D., 2004, ApJ, 609, 917

Maeder A., Mermilliod J. C., 1981, A&A, 93, 136

Manzi S., Randich S., deWitW. J., Palla F., 2008, A&A, 479, 141

Marsden S. C., Carter B. D., Donati J., 2009, MNRAS, 399, 888

Marsden S. C., Waite I. A., Carter B. D., Donati J., 2005, MNRAS, 359, 711

Mermilliod J. C., 1981, A&A, 97, 235

Meynet G., Mermilliod J., Maeder A., 1993, A&AS, 98, 477

Moraux E., Bouvier J., Clarke C., 2005, Astronomische Nachrichten, 326, 985

Oliveira J. M., Jeffries R. D., Devey C. R., Barrado y Navascu´es D., Naylor T., Stauffer J. R., Totten E. J., 2003, MNRAS, 342, 651

Pan X., Shao M., Kulkarni S. R., 2004, Nat, 427, 326

Parker Q. A., Phillipps S., PierceM. J., HartleyM., Hambly N. C., Read M. A., MacGillivray H. T., 15 co-authors 2005, MNRAS, 362, 689

Pinfield D. J., Dobbie P. D., Jameson R. F., Steele I. A., Jones H. R. A., Katsiyannis A. C., 2003, MNRAS, 342, 1241

Pinsonneault M., 1997, ARA&A, 35, 557

Randich S., Pallavicini R.,Meola G., Stauffer J. R., Balachandran S. C., 2001, A&A, 372, 862

Reid I. N., Hawley S. L., Gizis J. E., 1995, AJ, 110, 1838

Robichon N., Arenou F.,Mermilliod J.-C., Turon C., 1999, A&A, 345, 471

Sharp R., Saunders W., Smith G., Churilov V., Correll D., Dawson J., Farrel T., Frost G., Haynes R., Heald R., Lankshear A., Mayfield D.,Waller L.,Whittard D., 2006, in Ground-based and Airborne Instrumentation for Astronomy. Edited by McLean, Ian S.; Iye, Masanori. Proceedings of the SPIE, Volume 6269, pp. 62690G (2006). Vol. 6269 of Presented at the Society of Photo-Optical Instrumentation Engineers (SPIE) Conference, Performanceof AAOmega: the AAT multi-purpose fiber-fed spectrograph

Skrutskie M. F., Cutri R.M., Stiening R.,WeinbergM. D., Schneider S., Carpenter J. M., 25 co-authors 2006, AJ, 131, 1163

Stauffer J. R., Hamilton D., Probst R. G., 1994, AJ, 108, 155

Stauffer J. R., Hartmann L. W., Prosser C. F., Randich S., Balachandran S., Patten B. M., Simon T., Giampapa M., 1997, ApJ, 479, 776

Stauffer J. R., Schild R., Barrado y Navascues D., Backman D. E., Angelova A. M., Kirkpatrick J. D., Hambly N., Vanzi L., 1998a, ApJ, 504, 805

Stauffer J. R., Schultz G., Kirkpatrick J. D., 1998b, ApJL, 499, 219

van Leeuwen F., 2009, A&A, 497, 209

Whiteoak J. B., 1961, MNRAS, 123, 245

Williams K. A., Bolte M., Koester D., 2004, ApJL, 615, L49

Woo J., Demarque P., 2001, AJ, 122, 1602

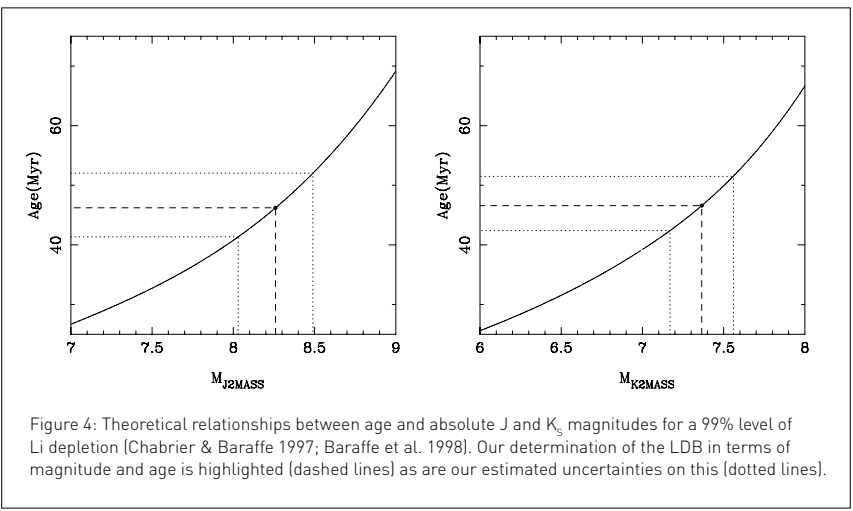

Figure 4: Theoretical relationships between age and absolute J and $K_s$ magnitudes for a 99% level of Li depletion (Chabrier & Baraffe 1997; Baraffe et al. 1998). Our determination of the LDB in terms of magnitude and age is highlighted (dashed lines) as are our estimated uncertainties on this (dotted lines).

# Making MANIFEST fibres for the Giant Magellan Telescope

Matthew Colless, Guy Monnet, Julia Tims, Will Saunders, Andrew Hopkins, Jon Lawrence, Jeroen Heijmans, Greg Smith, Lew Waller, Jurek Brzeski, James Gilbert, Michael Goodwin, Scott Smedley, Scott Case, Rolf Muller (AAO)

The AAO is proposing a concept for the Giant Magellan Telescope's Facility Multi-Object Fibre System called MANIFEST, the Many Instrument Fibre System. MANIFEST is a fibre-feed system for GMT that offers access to larger fields of view, higher multiplex gains, versatile reformatting of the focal plane via IFUs and image-slicers, and in some cases higher spatial and spectral resolution. Coupled to the seeing-limited instruments GMACS, G-CLEF and NIRMOS, it provides qualitative and quantitative gains over each instrument's standalone capabilities in terms of both instrumental parameter space coverage and scientific grasp and performance. The top right figure shows a schematic diagram of the main elements of MANIFEST.

MANIFEST is not an instrument per se, rather a generic telescope facility, like an adaptive optics adapter. Its Starbug system uses semi-autonomous probes hanging down under a glass field plate that hold and position optical fibres (see middle right figure). It allows hundreds of astronomical targets to be selected over the full GMT Gregorian focal plane, redirecting the light to the slit entrances of the 3 GMT seeing-limited instruments (NIRMOS, GMACS & G-CLEF).

"Starbugs" are discrete stepping devices using two co-axial piezoceramic tubes joined at their top ends to produce a 'walking' motion. The movement is analogous to a one-legged person operating a walking frame. Starbugs positioning is via a closed-loop control system with feedback via a high-resolution camera located above the field plate. Each Starbug has a set of back-illuminated fibres used as metrology markers. The asymmetric distribution of these fibres allows determination of the Starbug's orientation without movement or calibration routines. The prototype system with 10 Starbugs is shown in the bottom right figure.

**World-leading capabilities:** MANIFEST is not only a major force multiplier for GMT instruments and science; it also provides GMT with world-leading capabilities compared to the other ELTs. Although GMT is the ELT with the smallest aperture (A), it has the equal-largest field of view (Ω). It cannot beat the other ELTs at science where the figure of merit is the primary mirror diameter (D) to powers greater than unity (so-called $D^n$ science). However it *can* dominate wide field and survey astronomy (so-called AΩ science), because the other ELTs are not making use of the full field of view delivered by the telescope.

**Enhanced functionality:** MANIFEST offers GMT spectrographs a wide range of enhanced functionality: (1) increased fields of view; (2) multiple deployable IFUs; (3) increased resolution via image-slicing; (4) efficient detector packing, both spectrally and spatially; (5) efficiency gains from working at VPH super-blaze angles; (6) simultaneous observations with multiple instruments; (7) gravity-invariant spectrograph mounting; and (8) OH suppression in the NIR. Capabilities such as the deployable IFUs, increased multiplex and increased resolution enable entirely new science with GMT; other functionality offers efficiency gains ranging from incremental (e.g. efficient detector packing and working at the superblaze angle), to substantial (increased fields of view and simultaneous observations), to transformative (multi-IFUs and OH suppression).

**Versatility:** The MANIFEST concept increases the versatility of GMT by embodying three principles in its technical design – (1) *selectability*: instruments can work either in their native mode or with fibre feeds that reformat the focal plane in a variety of ways; (2) *modularity*: new instruments can easily be accommodated by adding new fibre modules; and (3) *upgradeability*: the system can provide new functionality via straightforward and relatively low cost upgrades of the fibre modules. MANIFEST is not an instrument but rather an integrated telescope facility, analogous to the adaptive optics systems in providing broad scientific utility and enhanced performance for a variety of instruments. MANIFEST expands the accessible parameter space and enhances the long-term potentialities of GMT's instrument suite.

The AAO has now completed the MANIFEST Feasibility Study for GMT and is currently carrying out a targeted R&D program while awaiting the outcome of the selection process for the GMT first-light instruments, a decision on which is expected early in 2012.

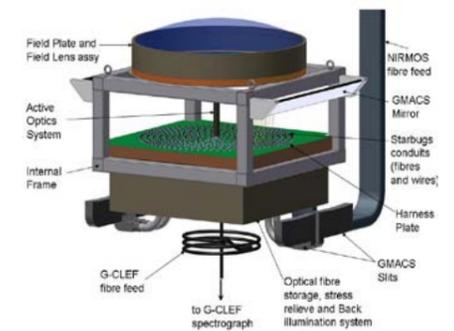

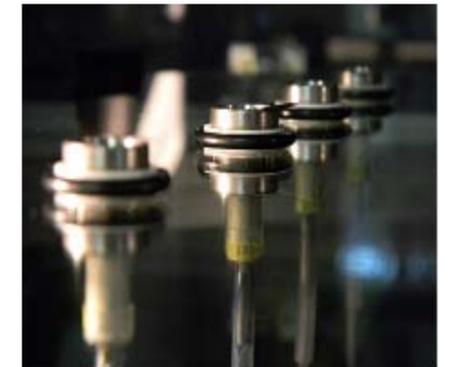

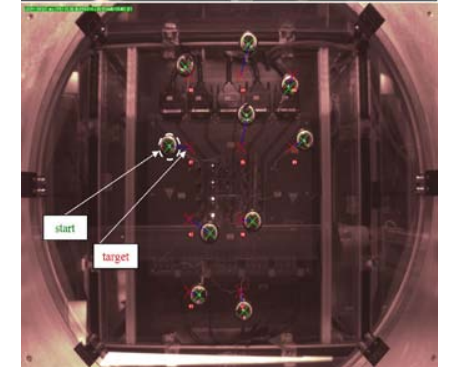

# A Voyage Through FOG

Kevin A. Pimbblet (Monash University)

Filaments of Galaxies (FOGs) have long been known about in the literature and discussed with a variety of mixed nomenclature (e.g. Walls, Sheets, Filaments), a famous early example being the CfA Great Wall (Geller & Huchra 1989) and more recently the Sloan Great Wall (Gott et al. 2005; Pimbblet et al. 2011). Over recent years, there has been increased interest in FOGs from the ready availability of large-volume redshift surveys N-body simulations (see Pimbblet 2005 for a general discussion).

My own interest in FOGs began during my PhD when I was undertaking observations of rich galaxy clusters using a new spectrograph that was called 2dF (e.g., Pimbblet et al. 2001). Fig. 1 displays a plot of one of these clusters (Abell 22; Pimbblet et al. 2005) where in front of the cluster (z~0.14), there appears to be coherent "wall" of galaxies in RA at lower redshift (z~0.11). This is clearly not the same mass regime or physical dimensions as the classic Great Wall from CfA. However, we find that it appears to connect up a set of distant galaxy cluster pairs together in non-trivial manner (Pimbblet et al. 2005). This opened up a large number of questions that included (but were not limited to): what role filaments have in the growth of clusters that they inter-connect? What effect do FOGs have on the evolution of galaxies contained therein? Can they be used as a cosmological probe to test theories of structure formation?

Using the final data release of the 2dF Galaxy Redshift Survey (Colless et al. 2003) coupled with a knowledge to the location of the galaxy clusters contained therein (De Propris et al. 2002), we undertook a systematic search for inter-cluster filaments (Pimbblet et al. 2004; PDH). PDH extracted all possible combinations of pairs of clusters within 10 degrees and $\Delta cz<1000$ km/s of each other. This generated a sample of 805 unique inter-cluster areas which were visually inspected for the presence of FOGs. Each filament that was confirmed in this manner was also given a morphological classification (straight, curved, wall or sheet-like, complex) and a length (explicitly, the inter-cluster axis regardless of morphology) was computed.

The relative fractions of morphological types are found to be entirely consistent with predictions from models (Colberg et al. 2005). However, the length of the filaments can stretch from to tens, to sometimes hundreds of Mpc which could pose a problem for models (cf. Yaryura et al. 2011).

If galaxies spend a lot of their lifespan inside FOGs, flowing toward clusters (see Pimbblet 2005), then presumably there could be some evolutionary pre-processing occurring inside FOGs before they ever encounter the hostile high-density cluster core regions. We know from Lewis et al. (2002) and Gomez et al. (2003) that star formation suppression starts to occur on a mass scale of galaxy groups. However those studies considered a radially averaged (or density averaged) star-formation rate. Hence the detail of what happens to galaxies inside filaments is «lost» in the averaging process. If we consider the star-formation rate as a function of distance away from a galaxy cluster, but only along a FOG vector, we find a result that was not seen in the previous analyses: a small blip (star burst) in star formation rate at a few Mpc away from cluster centres (Porter et al. 2008). We hypothesize that this starburst is caused by first time harassment.

FOGs should also contain most of the mass of the Universe (Cen & Ostriker 1999; Aragon-Calvo et al. 2010). Using archival data and a knowledge of where FOGs reside (i.e. PDH), we searched for X-ray emission from filaments and through a stacking process (carefully avoiding known contaminants such as other galaxy clusters) combined with a statistical background correction, derived an estimate for the electron density inside FOGs at z~0.1 $n_e = (4.7 \pm 0.2) \times 10^{-4}$ $h_{100}^{1/2}$ cm$^{-3}$ (Fraser-McKelvie et al. 2011). This is, perhaps, a higher than expected density and may reflect the stacking & background subtraction process and/or the modelling of the X-ray count rate. Future work with new satellites such as Suzaku should clarify this preliminary estimate.

This article necessarily represents a very quick whip-around of our results, largely derived from AAO data products—particularly redshift surveys conducted with 2dF and AAOmega. It is a pleasure to thank the dedicated staff of the observatory and colleagues in large-field surveys that made observational research in to FOGs a very promising avenue of investigation.

## REFERENCES

Aragon-Calvo M. A., van de Weygaert R., Jones B. J. T., 2010, MNRAS, 408, 2163

Cen R., Ostriker J. P., 1999, ApJ, 514, 1

Colberg J. M., Krughoff K. S., Connolly A. J., 2005, MNRAS, 359, 272

Colless M., et al., 2003, arXiv:astro-ph/0306581

De Propris R., et al., 2002, MNRAS, 329, 87

Fraser-McKelvie A., Pimbblet K. A., Lazendic J. S., 2011, MNRAS, in press

Geller M. J., Huchra J. P., 1989, Sci, 246, 897

Gomez P. L., et al., 2003, ApJ, 584, 210

Gott J. R., III, Juric M., Schlegel D., Hoyle F., Vogeley M., Tegmark M., Bahcall N., Brinkmann J., 2005, ApJ, 624, 463

Lewis I., et al., 2002, MNRAS, 334, 673

Pimbblet K. A., Smail I., Edge A. C., Couch W. J., O'Hely E., Zabludoff A. I., 2001, MNRAS, 327, 588

Pimbblet K. A., Drinkwater M. J., Hawkrigg M. C., 2004, MNRAS, 354, L61

Pimbblet K. A., Edge A. C., Couch W. J., 2005, MNRAS, 357, L45

Pimbblet K. A., 2005, PASA, 22, 136

Pimbblet K. A., Andernach H., Fishlock C. K., Roseboom I. G., Owers M. S., 2011, MNRAS, 410, 1837

Porter S. C., Raychaudhury S., Pimbblet K. A., Drinkwater M. J., 2008, MNRAS, 388, 1152

Yaryura C. Y., Baugh C. M., Angulo R. E., 2011, MNRAS, 413, 1311

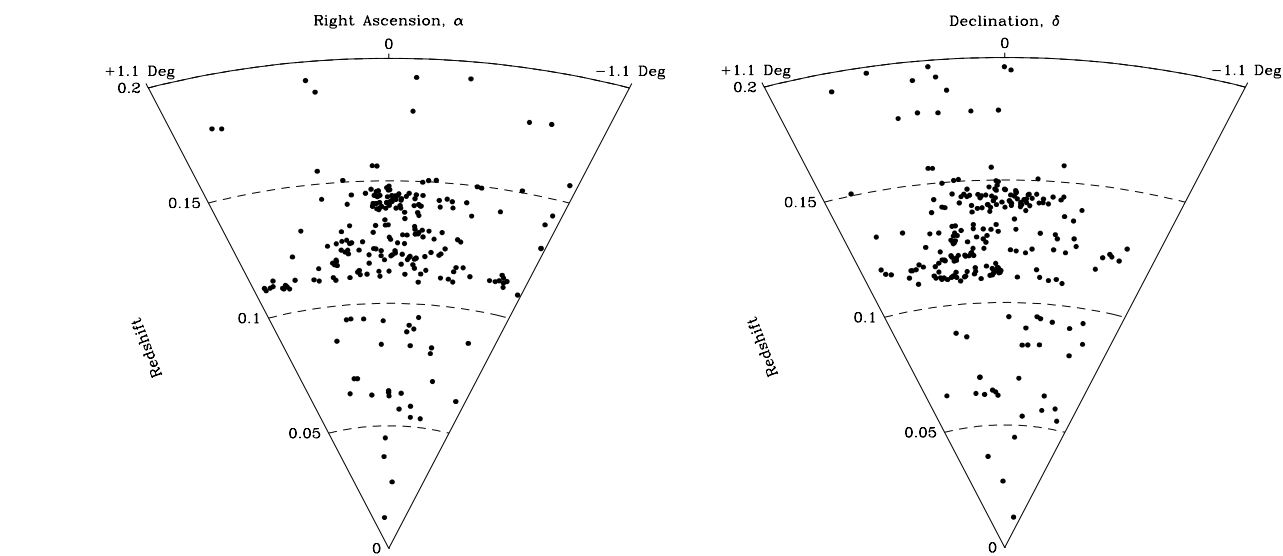

Figure 1: Wedge plots of right ascension and declination versus redshift in the direction of Abell 22. The X-ray centre of the cluster is located at $\alpha = \delta = 0$. Note the ensemble of galaxies in the foreground of the cluster. Although one may expect this plot to exhibit a 'finger of god' effect for the cluster, arising from the distorting effects in redshift space, this is not seen, as the elongation in redshift space of the cluster is small in comparison to the depth of redshift space covered.

# Are You Biased?[1]

Sarah Brough (AAO), Tanya Hill (Melbourne Planetarium), Amanda E. Bauer (AAO), Andrew Hopkins (AAO), Sarah Maddison (Swinburne University)

Picture this: you are running late for a meeting. You race down the corridor and there are two rooms to choose from. One is full of blokes vigorously debating something, the other contains a group of women animatedly discussing something else. Which door would you choose?

It's a bit of a cliché, and there will be plenty of exceptions, but, regardless of the subject being discussed, a good many of us would simply choose the door that contains those of the same gender as us. Even if the conversation bores you crazy, at least among members of your own gender, you can mostly disappear into the background without too much fuss.

If you feel ostracised by these comments – well that's the point isn't it? It's a basic human desire to feel like we belong. But can you recognise your **unconscious biases**? Do you know how this affects your interactions with people, especially in the workplace?

This article follows from a recent "Women in Astronomy" workshop organised by the Women in Astronomy Chapter of the Astronomical Society of Australia (ASA) and sponsored by the AAO, as well as the Centre for All-sky ASTROphsyics (CAASTRO) and CSIRO Astrophysics and Space Science (CASS). The aim of the workshop was to investigate why many women leave astronomy at the mid-career level but the conclusions we drew have positive implications for many members of the astronomy community.

## Unconscious Bias

Why is unconscious bias relevant? In a non-gender balanced field like astronomy, historically driven by senior men, unconscious bias leads to non-diversification which loses us the variety of opinion and thought processes that are vital to research excellence. Unconscious bias also means that people who are not part of the dominant group feel ostracised and that their contributions are not respected.

[1] This article is based on the Women in Astronomy Workshop 2011 Report, available at http://asawomeninastronomy.org/meetings/wia2011/

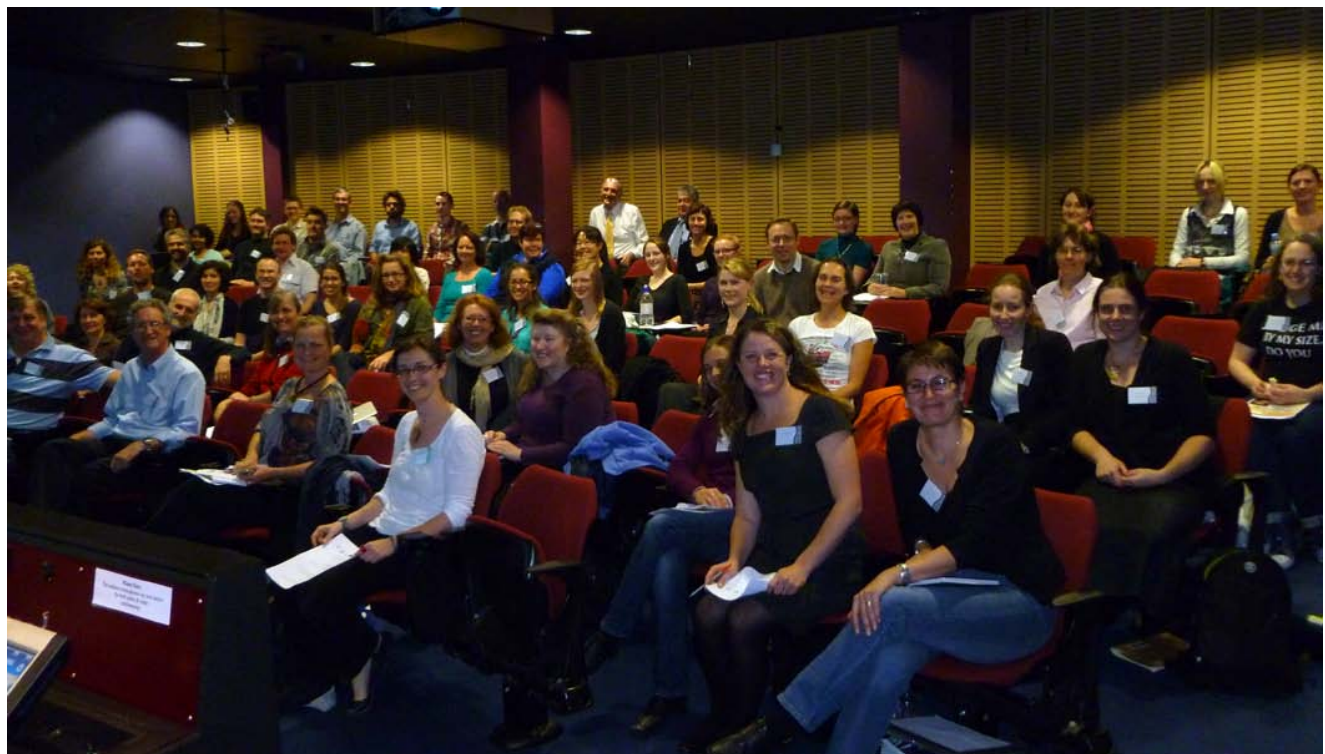

Attendees at the recent Women in Astronomy Workshop. Image credit: Bryan Gaensler

We all find it easier to work with people with whom we have something in common. On this basic level, a male manager (and they're in the majority) might choose to give a big job to a younger male colleague because he remembers the conversation they had after a meeting about a shared interest in footy, for example. Recognising that you might have a tendency do this is a good start. Then trying 50% of the time to go outside your comfort sphere and offer the job to someone else, who might be just as deserving but not having an obvious common interest, will impact positively not only on the diversity of people working in astronomy, but the happiness of those people. Everyone wants to be valued.

It is not clear whether the lack of women on committees, in invited talks, and appointment panels is forgetfulness or unconscious bias, but it certainly does not help address the problem of diversity. Not only do these people make decisions for the future of our community, the invitations, awards and committee membership also confer community respect, necessary for grant applications amongst much else. Women may not spring to mind immediately for these positions, but they are there. The more exposure given to women, the more they will spring to mind, equally.

There are unwritten rules for how we judge people based on gender. Men and women will judge a strong man as successful and want to work for them, while a strong woman is often seen to be aggressive and not someone you would want to follow (Steinpreis et al. 1999). When reading or writing reference letters it is important to think carefully about the adjectives used. Madera et al. (2009) found that it is common for women to be described in nurturing terms (happy, contributes …) and men in strong terms (go-getter, confident … ) where generally it is the strong terms that are sought after by selection committees.

This all sounds very negative but it is relatively easy to fix. Firstly, we all need to be aware of our personal biases, and think carefully before selecting committee members, recipients of awards, invited talks or writing reference letters. Secondly, by stacking our committees with women. Trying to achieve 50% balance in committees, appointment panels, recipients of awards and numbers of invited talks will bring different views into play as decisions are made for astronomy's future, encouraging future diversity.

### Family vs Career

This is a major cause of women leaving astronomy at every level. For those of you who have not noticed, there is a neat correlation between the age when people are beginning to get ongoing positions and the biological deadline when women need to think very seriously if they have not had children yet. Having children means taking time off and being less research productive for a period of time, making it even harder to get that elusive ongoing position. As a community we need to recognize the existence of this correlation and that we must have the flexibility and foresight to allow astronomers the opportunity to have a career and a family. Institutions must not be penalised for having flexibility and foresight and no colleague should have to choose "Family vs Career".

For those researchers on fixed-term contracts, if their contracts are increased by the same length of time as the parental leave taken, researchers can get their research done and institutions can get the research they employed someone to do!

CSIRO (amongst others) has also shown that flexible grants designed to encourage parents to return to research after parental leave are incredibly valuable, not just to the recipients but the research community around them. Several recent workshops at CSIRO Astrophysics and Space Science have been fully or partially funded by Ruby Payne-Scott awards.

For employees with children, there are things that institutions can do to make themselves Family Friendly. They can make sure that important meetings are held between 10am and 2pm, and that meetings finish on time, as this helps everyone be able to engage with their families. Surely that is their most important role in life?! Institutions also need to think seriously about ensuring the availability of childcare close to work and at professional meetings, especially national meetings. The closer/cheaper/easier the childcare, the more researchers will return to work quicker and happier. Childcare at professional meetings means that more primary caregivers can participate in those meetings they may not have been able to attend otherwise.

As we continue to peer-review each other's grant, award or job applications, we need to actively take into account career breaks (of all varieties), and non-100% research roles, when taking responsibility for such judgments. Also, in assessing candidates for job/grant applications, we should be asking for best N years of publications, rather than previous N years, to account for those career breaks.

Amazing researchers even in recent times have had to "hide" the fact that they have kids, have taken time off, or that they work part-time, all to compete. For researchers of this generation, our challenge is to make all these normal and open to acceptance. Not hidden away.

### Imposter syndrome

Despite external evidence of their competence, individuals with imposter syndrome remain convinced that they are frauds and do not deserve the success they have achieved. Proof of personal success is dismissed as luck, timing, or as a result of deceiving others into thinking they are more intelligent and competent than they believe themselves to be. Imposter syndrome affects everyone. As individuals we need to face this, stand up to it and then be empowered by overcoming it! Managers need to be aware that their staff may suffer from this and that they **can** help! For whatever reason, some people need a champion to help them apply for jobs, promotions, or to ask a question in a meeting. A good manager can do this but additional mentoring is **crucial**. Having a senior member of the community (male or female) providing advice and encouragement is vital no matter what career stage you are at. Anything we can do, as a community to help people forge the links that result in mentoring is essential. If your institution is small, there may be suitable external mentors from collaborative organisations, and there are companies who can provide mentoring.

### Handy Hints for Individuals

Individuals need to recognise that what you say/do does not have to be perfect. Do not think you need to know everything and be 110% precise before you speak up. Your vocal contribution of substance **IS** valuable and **IS** necessary for you to be noticed, heard, acknowledged and appreciated.

Recognise when you are being handed a task that will not benefit your career, anecdotally given to women and hence called **pink** tasks. These are tasks that need to be done on time and to a high standard, but where there is little substantive development or increased visibility for the person assigned the task. While these tasks are necessary, when they are assigned to the same person over and over again, that person is unlikely to advance. If you find yourself repeatedly taking these tasks that do not drive your career forwards, you need to say no. The point is not to be liked, but to be respected.

An ongoing theme in the workshop was that people who had asked for dispensation or consideration or a special case **received** it! Bear that in mind: If you want something, you will not get it without asking, so Ask! You never know what will happen if you ask.

### Summary

Very few career paths are actually 'traditional' and they do not need to be. Take it for what it is. We should not need to contort ourselves to fit the 'traditional' path and should never accept being told otherwise. Whilst enjoying our varied careers, we need to accept the differences our colleagues bring to the table and use our understanding of the positive benefits of such diversity to help us work together. Which is really what this article is all about - creating environments where we **all** feel we belong, can contribute and will be respected.

### References

Madera, J. M., Hebl, M. R., Martin, R. C. (2009). "Gender and Letters of Recommendation for Academia: Agentic and Communal Differences", J. Applied Psychology, 94, 1591

Steinpreis, R. E., Anders, K. A., Ritzke D. (1999). "The Impact of Gender on the Review of the Curricula Vitae of Job Applicants and Tenure Candidates: A National Empirical Study", Sex Roles, 41, 718

# AUSGO Corner

Stuart Ryder (Australian Gemini Office, AAO)

## Proposal Statistics

For Semester 2011B ATAC received a total of 34 Gemini proposals, of which 21 were for time on Gemini North, 4 were for exchange time on Keck or Subaru, and 9 were for time on Gemini South. The overall oversubscription (1.81) was up on 2011A, although Gemini South was barely oversubscribed. At the ITAC meeting Australia was able to schedule 24 programs into Bands 1–3 (half of which involved joint allocations with other Gemini partners), and one more in the expanded Poor Weather Queue. For Magellan we received 12 proposals, resulting in the highest Magellan oversubscription ever of 4.86, due in part to the availability of the new wide-field near-IR camera FourStar.

In 2010B, all but one of the seven Band 1 programs were completed or had insufficient Target of Opportunity (ToO) triggers; 4 of 6 Band 2 programs got 90% or more of their data or did not trigger ToOs; while no Band 3 programs were even started. While Australia used barely half its allocated time in 2010B, nearly 80% of that time went to programs that were completed.

## Magellan and Gemini travel funding

The ANSTO-administered Access to Major Research Facilities Program (AMRFP) which has supported travel to overseas observatories for a number of years finished in June 2011. AusGO and AAL approached DIISR with a request to use some uncommitted travel funds from the National Collaborative Research Infrastructure Strategy (NCRIS) to cover Magellan observer travel in Semester 2011B, which DIISR approved as a one-off gesture and not a precedent. However with NCRIS itself also finishing in 2011, the AAO and AAL are actively exploring options to ensure that users who are fortunate to be awarded classical time on Magellan, Gemini, or Gemini exchange time on Subaru or Keck are actually able to take up that opportunity.

## Gemini High-resolution Optical Spectrograph

On 6 May 2011, the Gemini Observatory issued a Request for Proposals for Conceptual Design Studies for a Gemini High-resolution Optical Spectrograph (GHOS). A team led by Dr Michael Ireland from AAO/Macquarie University with involvement from AAO and RSAA, as well as KiwiStar Optics in New Zealand has submitted a bid for a concept design study, building upon the AAO's experience with the CYCLOPS image slicer and fibre-feed for UCLES, as well as HERMES and the multi-object fibre positioner MANIFEST for GMT.

## NIRI Replacement

With the commencement of the GHOS procurement, the Gemini Science Committee (GSC) is now turning its attention to the next Gemini instrument. With concern growing about the reliability of NIRI, the GSC asked all National Gemini Offices to poll their user communities on their future needs and desired capabilities for near-infrared imaging on Gemini North. AusGO ran a survey for the Australian community, with results forwarded to the GSC on 8 June. Our findings were that: (a) only 20% support an L' & M band capability for a replacement instrument; (b) two-thirds see no need for a spectroscopic or other additional modes in any replacement; (c) about half plan to use NIRI in "wide-field" (2') or Adaptive Optics mode in coming years; (d) 87% are fully or mostly satisfied with current NIRI performance; and (e) about half feel that NICI could substitute for NIRI if need be.

## GeMS First Light

The Gemini Multi-Conjugate Adaptive Optics System (GeMS) consisting of the Canopus optical bench; a 50 W laser to produce a "constellation" of 5 laser guide stars over an 85" field; and the RSAA-built Gemini South Adaptive Optics Imager (GSAOI) has been undergoing commissioning in 5 night blocks each month throughout Semester 2011A. While initial results are promising, much work remains to be done to tune the full system and the many control loops which must operate to correct for atmospheric turbulence across the full field of view (unlike ALTAIR on Gemini North, for which the correction deteriorates going radially outward from the natural or laser guide star). The image shown here is an "engineering first light" image from GeMS taken on 19 April 2011 and demonstrates that already the system can deliver remarkably uniform images across most of the GSAOI field. A call for System Verification observations with GeMS is likely to be issued during Semester 2012B.

## Gemini School Astronomy Contest

Building on the success of the 2009 (IYA) and 2010 Gemini School Astronomy Contests, AusGO is running a similar contest in 2011 for high school students to win one hour of time with GMOS-South allocated by ATAC in 2011A. The number of entries was up by 20% on 2010. The winning entry was submitted by Benjamin Reynolds from the Sutherland Shire Christian School (NSW), to observe the barred spiral galaxy NGC 7552 with its nuclear star-forming ring. The two runners-up were Ryan Soars from Trinity College (WA) who suggested the YSO NGC 6729, and students from St Margaret's Anglican Girls School (QLD) who suggested "Burbidge's Chain" of galaxies. Observations for the winning entry are now in the Gemini South queue awaiting observation.

The images for the two previous winners have been incorporated into A3-sized posters promoting the contest, Gemini, the AAO, and AAL. These were distributed at the ASA Annual Scientific Meeting in Adelaide and are freely available for outreach purposes.

## Gemini Town Hall meeting

As part of the Gemini Observatory's effort to engage with their user community on a regular basis, Associate Director for Operations Dr Andy Adamson followed up his visit to Australia last September with a Gemini "Town Hall" meeting held in conjunction with the ASA ASM in Adelaide. He provided an update on the status of the Gemini Observatory, including recent changes in management and a review of developments in the "transition plan" for the observatory as they move into the final year of UK membership. The instrument development program was also reviewed, followed by a Q&A session afterwards.

## Observational Techniques workshop

From 30 Aug–2 Sep 2011 the AusGO and AAT Science Workgroups will run an optical/IR observational techniques workshop in Sydney. This is the first time such a workshop has been run by the AAO for a decade; the early level of interest in attendance by students and even postdocs highlights that such a workshop is long overdue. Presentations on all aspects of imaging, photometry, and spectroscopy available on the wide variety of instrumentation offered on the AAT, Gemini, and Magellan will be given primarily by AAO and Gemini Observatory staff, together with data reduction tutorials. Details of the workshop are available at http://www.aao.gov.au/AAO/AUSGO-AAO_Workshop/

## Gemini e-newscast

To help stay in touch with recent developments, announcements, and news releases, why not subscribe to Gemini's e-newscast service? To subscribe to the list, send a message to listserver@gemini.edu with a subject of "subscribe Gemini-eNewscast" (without the quotation marks). Previous issues can be accessed at http://www.gemini.edu/enewscast.

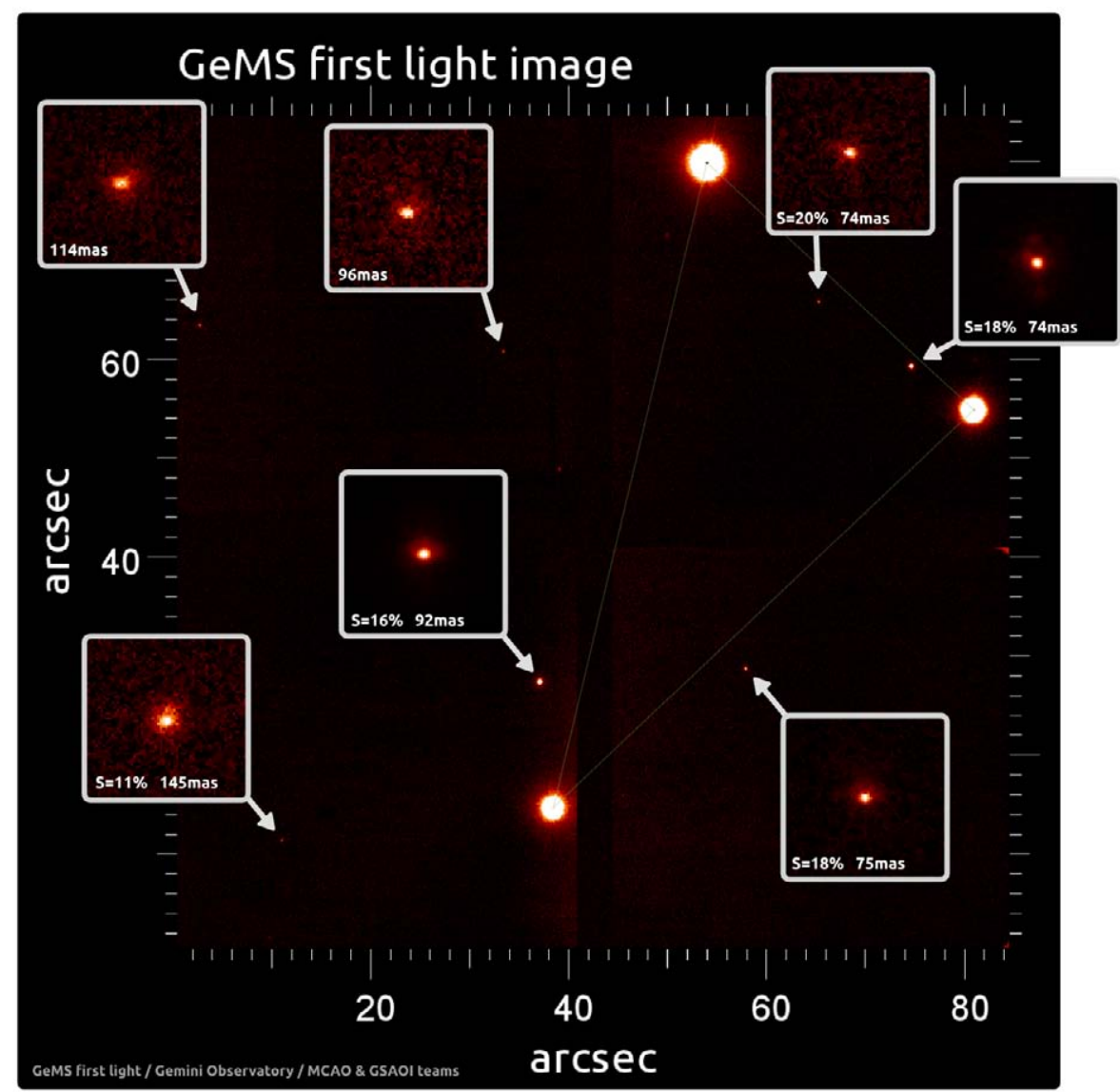

Figure 1: "First light" image obtained with GeMS and GSAOI on Gemini South at a wavelength of 2.12 microns. Strehl ratio and full-width at half-maximum values for all stars are shown in the insets. The poorer image quality on the left edge is expected, as these stars are outside the constellation defined by the three bright stars in the right half used to provide the tip-tilt correction.

# Supernovae and their Host Galaxies
## Sydney, Australia, 20-24 June 2011

Chris Lidman (AAO)

During a marvellously sunny week at the end of June (as can be verified by the group photo), 150 astronomers from all parts of the world came to Sydney to talk about supernovae and their host galaxies. The week long conference was held at the Australian National Maritime Museum in Darling Harbour, and was the fourth conference in the Southern Cross Astrophysics Conference Series, which is jointly sponsored and organised by the AAO and CASS.

The conference started with a public talk given by Prof. Robert Kirshner at the Powerhouse Museum. Prof. Kirshner captivated the audience with his description how astronomers uncovered what is now considered one of the biggest unsolved mysteries of physics - the reason why the expansion of the Universe is accelerating. His talk also reminded some of us that doing science is fun.

A wide range of topics were covered at the conference: from the progenitors of core-collapse and thermonuclear supernovae to the influence that supernovae have on the properties of their hosts and visa-versa. Of special interest to the author of these lines are the new types of supernovae being discovered by the current wide-field transient surveys. We also heard about future new surveys for transients, including surveys for transients at radio wavelengths. There was also a special session on SN 2011dh, a supernova that was discovered in the Whirlpool Galaxy (M51) just a couple of weeks before the start of the conference.

Even though the scientific program was very active, there was time for social activities. Brian Schmidt hosted a wine tasting event at the Museum, where we were able to sample a range of wines, including one of his own, from Australia and New Zealand, we got to see Saturn through telescopes at the Sydney Observatory that were probably older than any of the participants, and we enjoyed some nice views and nice food at the conference dinner at Doltone House.

The conference would not have been possible without the dedicated work of the local and scientific organising committees, the members of which are listed below, and to the many people who helped at the conference: Robert Barone-Nugent, Erica Rosenblum, Philippa Morley, Janine Myszka, Stacey-Jo Dyas, Billy Robbins, Kitty Lo, Jason Spyromilio, and Amanda Bauer. On behalf of the participants, I thank all of you. A very special word of thanks goes out to Vanessa Bugueno, Angel Lopez-Sanchez, and Paul Hancock for the great work that they did before, during and after the conference.

**The Scientific Organising Committee:**

Alex Filippenko, Claes Fransson, Bryan Gaensler, Andy Howell (co-chair), Rubina Kotak, Chris Lidman (co-chair), Filippo Mannucci, Seppo Mattila, Ken'ichi Nomoto, Giuliano Pignata, Brian Schmidt, Alicia Soderberg

**The Local Organising Committee:**

Robert Braun, Ricardo Covarrubias, Paul Dobbie, Andrew Hopkins, Chris Lidman (chair), Angel R. Lopez-Sanchez, Tara Murphy, Quentin Parker, Stuart Ryder, Helen Sim, Max Spolaor

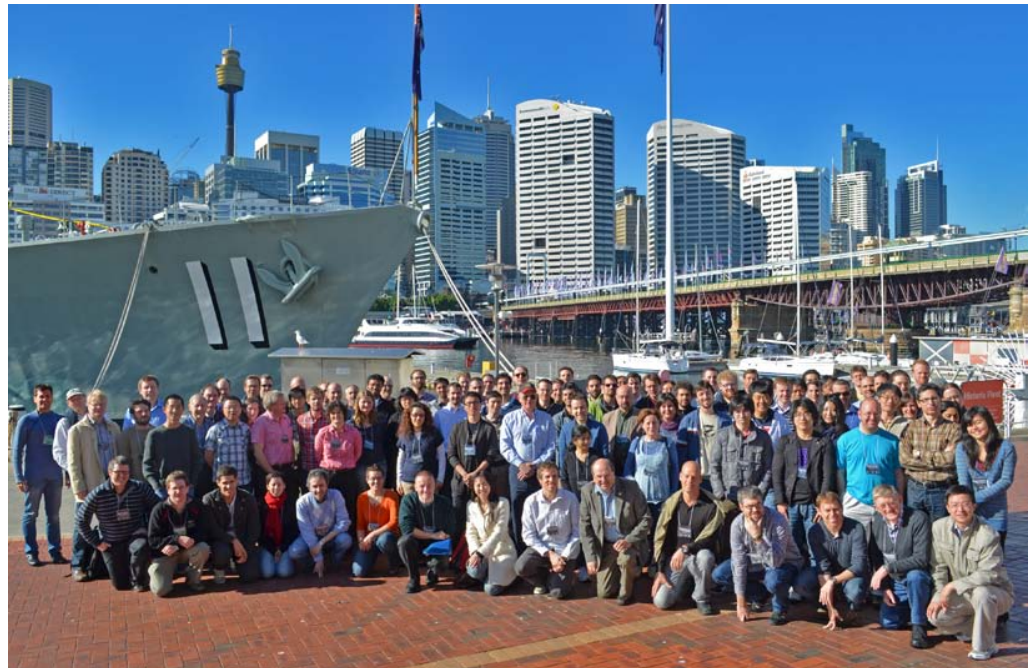

Conference Group Photo

# Scientist Meets Parliament

Sarah Brough (AAO)

I was invited to attend the 2011 Science Meets Parliament in Canberra in June this year as the representative of the Astronomical Society of Australia (ASA), along with several other astronomers from around Australia. The event is organised by Science & Technology Australia (formerly the Federation of Australian Scientific and Technological Societies; FASTS) to give scientists the opportunity to meet Federal politicians and discuss science issues with them. In preparation for that meeting we also had a day of training in communicating science to the media, general public and politicians.

The event started with the launch of the 'Respect the Science' campaign to explain to the general public how the scientific and peer review processes weight information from scientists, against those whose words have not gone through those processes. Followed by a presentation on the current political climate, informing us that it is very, very local, and very hard to get a mention in between the current 'BIG' issues.

We had a 'Meet the Press' session with political reporters from the parliamentary press gallery helping us understand how to get our science into the general media. Some of the messages I took away were, if you do have an interesting press release, send it out before the release day to a targeted journalist so that they have the 'hint of exclusivity' as that will help you get coverage. Also that pictures are appreciated, as well as, if possible, case studies to humanise the story (although I'm not sure this is applicable to astronomy).

After lunch each person prepared and presented a 45s piece on our research (in less than the time it takes a sparkler to burn). This turned out to be a very valuable exercise prior to our meeting the politician on the second day.

In the afternoon we had presentations on how to convey our message to politicians. We were informed to leave plenty of time for small talk, to be able to think up good stories with our science, and come up with verbal pictures to connect large numbers to reality, to go in to have a conversation in their language (emphasising what is important to them) and form a relationship.

We were advised not to lecture, information dump, whinge about funding, offer mixed messages or have nothing to say. Also that we should be careful about how uncertainty is communicated, as 'almost certain' is not really understood. It was emphasised that preparation on our part is key - to research our assigned minister and their interests - and that "what excites a teenager will excite a parliamentarian".

The end of Day 1 was the dinner in Parliament House. Sadly several politicians were unable to attend so there was no parliamentarian at our table. Over dinner Former Victorian Premier John Brumby, gave an overview of science funding during his years in the Victorian Labor government. Followed by FASTS President Cathy Foley introducing the 'Respect the Science' video and the new name for FASTS: Science & Technology, Australia.

The second day started early with a breakfast briefing on the Excellence in Research for Australia (ERA) process from Professor Margaret Sheil, CEO of the Australian Research Council. It was interesting to hear the reasoning for having the ERA process, the acknowledgement that the choice of assessment metric drives behaviour so has to be thought through, as well as some of the outcomes and the improvements that are being implemented for the next round in 2012.

Following the breakfast briefing the imminent closure of Canberra airport due to the Chilean volcano ash cloud caused a brief hiatus as everyone tried to make new arrangements for travel and accommodation.

Our second presentation of the day was from Senator Kim Carr, the Minister for Innovation, Industry, Science and Research. He gave a very good, impassioned speech about the value of science evidence-based policy and the new 'Respect the Science' campaign.

We were fortunate to be able to attend new Chief Scientist Professor Ian Chubb's first National Press Club Address. This was fascinating, as well as being the best meal of the two days. Professor Chubb reiterated the 'Respect the Science' and peer review process message that underpinned this meeting. He also showed his experience in expertly handling questions from journalists. It is clear that he will take the science agenda to parliament.

From the Press Club we sat in the Public Gallery of the National House of Representatives to watch Question Time. This was also fascinating with The Speaker of the House introducing the Science Meets Parliament delegates to the House, resulting in a wave from Prime Minister Julia Gillard (we all waved back).

My meeting with Senator Chris Evans, Minister for Tertiary Education, Skills, Jobs and Workplace Relations, was scheduled shortly after the formal end of the day. John O'Byrne (University of Sydney), two other scientists and I went to meet the Senator. As he was still speaking in the Senate we ended up meeting with one of his advisors, Andrew Dempster. Luckily I was well-prepared by the previous day's presentations and I spoke quickly on the value of Square Kilometre Array (SKA) to galaxy research and broader issues such as enthusiasm in science, jobs and technological skills with processing of such huge quantities of data. I was impressed by Mr Dempster's awareness of the SKA.

I greatly enjoyed myself at Science Meets Parliament and found meeting so many other scientists very valuable. I would like to thank the ASA for sponsoring me to attend and Science & Technology Australia for organising the event. I would definitely recommend it to anyone interested in science communication or the parliamentary process.

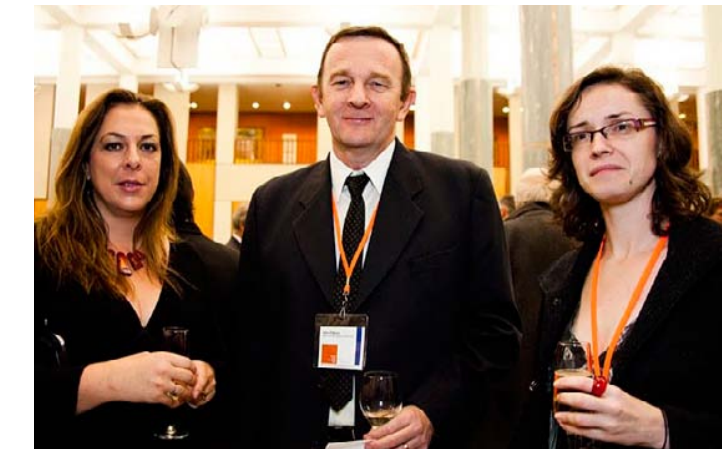

Foyer of new Parliament House before Dinner - John O'Byrne (centre), Sarah Brough (right). Image credit: Lorna Sim / Science & Technology Australia

# Epping News
Sarah Brough (AAO)

Two members of AAO staff have won annual prizes announced by the Astronomical Society of Australia (ASA) recently:

**Gayandhi De Silva** has won the Louise Webster prize. This prize is awarded in recognition of outstanding research by a scientist early in their post-doctoral career. Gayandhi's award was based on the research she published establishing the viability of the chemical tagging technique and setting the stage for the field of Galactic Archaeology: Detailed chemical abundance patterns of stars offer the possibility to reconstruct some components of the protogalactic disk and so improve our understanding of the Galactic disk formation process. If star-forming aggregates can preserve unique chemical signatures within its member stars, we can use this signature to tag dispersed individual stars to a common formation site. This is the concept of chemical tagging.

De Silva et al. 2007 (AJ, 133, 694) was the first to demonstrate the viability of the chemical tagging technique. Using UCLES/ AAT data they derived abundances for elements of various nucleosynthesis origins for stars in the HR 1614 moving group. They discovered these dispersed stars were metal rich with [Fe/H] > 0.25 dex, and that the star-to-star abundance scatter was less than 0.03 dex across all elements. This was the first time that stars which are physically dispersed and located all over the sky, were found to have near identical abundances and kinematics. Hence this was the first identification of an ancient relic star forming event in the disk and paved the way for large scale applications of the chemical tagging technique.

**Max Spolaor** has won the Charlene Heisler prize. This prize is awarded for the most outstanding PhD thesis in astronomy or a closely related field, accepted by an Australian university. The thesis research must show outstanding excellence and originality. Max's thesis produced important results relating the metallicity gradient and host galaxy mass for early-type galaxies. At least 5 first-author refereed publications resulted directly from his thesis, which was also awarded an internal prize for best PhD thesis at Swinburne University.

We congratulate them both!

**Sarah Brough** (and that's me so I will continue in the first person) - I have accepted a new position within the AAO as a Research Astronomer for the next five years. I will be continuing and developing my research with the Galaxy And Mass Assembly (GAMA) survey, Brightest Cluster Galaxies and Integral Field Spectroscopy and hoping to combine the three in the near future. I will also be taking on the AAOmega Instrument Scientist role and seeing through the planned AAOmega refurbishment.

We have two new faces in the Mechanical Group in Epping: **Nicholas Staszak** is the New Mechanical Manager and started in March and **Naveen Pai** has joined us as a mechanical technician, in which role he will be taking over many of the duties currently undertaken by **Denis Whittard**.

We have also enjoyed a morning tea to celebrate everyone who has been at the observatory and/or in the public service for more than 20 years: **Helen Davies, Tony Farrell, Garry Kitley, Ed Penny, Keith Shortridge, Lew Waller, and Helen Woods** (who spent 13 years in the foreign service with DFAT) got certificates. More recently **Don Mayfield** joined this list of luminaries.

Sadly people have also left the AAO. We wish them all the best in their new endeavours!

**Ricardo Covarrubias** was one of the first Magellan Fellows. He was initially seconded to the Carnegie Observatories for 2 years supporting visiting Magellan observers at Las Campanas Observatory, followed by a 6 month extension. In his time as a Magellan Fellow, Ricardo set new standards for observer support at Magellan. He spent his last 15 months here at the AAO working on various aspects of supernova research as well as serving on the LOC of the Southern Cross Astrophysics Conference on "Supernovae and their Host Galaxies", held in June.

In May, **Guy Monnet** retired from his position as acting Head of Instrumentation at the AAO following Sam Barden's departure. We wish him a quiet retirement from now on.

After 10 years at the AAO, **Scott Smedley** has decided to move on and undertake a new challenge. For the next phase of his career, he is going to be using his Linux/C++ skills developing trading systems in the Finance world. This will surely be very different to his past experiences!

We farewell-ed **Denis Whittard** who will be retiring, again, from the AAO after 23 years of service.

**Max Spolaor** is leaving after just a year as an AAO Research Fellow having taken on the AAOmega Instrument Scientist role and been responsible for setting in motion refurbishment plans for the AAOmega instrument. Max has been awarded a NASA Postdoctoral fellowship based at the University of California, Los Angeles.

## Summer Students

The AAO runs a twice-yearly fellowship programme to enable undergraduate students to gain 10-12 weeks first-hand experience of astronomical-related research. The current crop of students are from all around the world, enjoying spending our winter learning new skills.

**Stacey-Jo Dyas** is working under the supervision of Gayandhi de Silva on a chemical analysis of the open cluster, IC 4756. Such a study allows for the investigation of many processes occurring within the cluster; mixing mechanisms in stellar envelopes are one example. In addition, this type of research is crucial for the success of HERMES. HERMES aims to unravel the formation history of the Milky Way by identifying stars that formed from the monolithic collapse of a single gas cloud and tracing their paths through time, back to an earlier, common position. In order to identify members of the same clusters it is necessary to understand the key chemical signatures that cluster members share.

Stacey-Jo comes from the United Kingdom and is currently transitioning between institutes. Her undergraduate studies were at University College London where she recently obtained a Natural Science MSci. Stacey-Jo's final year project investigated the impact of image quality on cosmological measurements for the Dark Energy Survey. In October, she begins work on her doctorate at the Institute of Astronomy, University of Cambridge. Stacey-Jo has previously held summer research placements with the Royal Astronomical Society (2010) and the European Southern Observatory (2009).

**Janine Myszka** has just finished her third year as an undergraduate at Villanova University in Philadelphia, Pennsylvania, where she is currently studying for her B.S. in Astronomy & Astrophysics. She is working under the tutelage of Dr. Paul Dobbie and Dr. Chris Lidman to identify white dwarfs in the deep field images from the Canada France Hawaii Telescope Legacy Survey. These may ultimately serve as high fidelity calibrators for the studies which use these datasets. Janine enjoys observing the Coriolis effect in her everyday life and hopes she doesn't meet a funnel web spider during her stay.

**Erica Rosenblum**, grew up and went to school in New York State, USA. She recently received her bachelor's degree in Physics from Stony Brook University and is looking forward to working in New York City in the fall. Here at the AAO, Erica is working with Dr. Maritza Lara-Lopez on a project involving the relationships between metallicity, star formation rate and stellar mass in galaxy pairs using the GAMA (Galaxy and Mass Assembly) Database.

**Faustine Cantalloube** is an optics engineer masters student at the "Institut d'Optique, Graduate school" in France (in partnership with the Paris-11 University, Orsay). Faustine is working at the AAO for eleven weeks in the astronomical instrumentation field. Her project is to build an experimental set-up which could verify the liquid atmospheric dispersion compensator (LADC) concept. Faustine will therefore investigate the subject and performing simulations with the Zemax software to check the consistency of her work. She will also build and calibrate a compact optical layout in the AAO laboratory. This will enable future tests of the potential chemicals that can play the role of a prism.

## Vale Tom Cragg (1927-2011)
Steve Lee (AAO)

I've always thought that Tom Cragg could best be described as a professional amateur astronomer. Throughout most of his working life he had been able to mix his love of astronomy with earning a living. Tom was Chief Night Assistant at the AAT from 1976 until his retirement in 1992. Hired during the early years of the AAT to bring some experience to the new telescope, Tom came from Mt. Wilson Observatory where he had worked for the previous 24 years.

Born in 1927 in St. Louis, Missouri his family moved to Los Angeles when he was 8. Like many of us, he formed an interest in astronomy during his early school years and never let it go. He joined the LA Astronomical Society, and also became a volunteer at the Griffith Planetarium, showing visitors the wonders of the night sky. A chance encounter with renown solar astronomers Seth Nicholson and Robert Richardson while he was visiting Mt Wilson Observatory led him to him working at the Pasadena office reducing solar observations, and eventually to becoming resident solar observer at the 150-foot solar tower on Mt. Wilson - a position he held for the 14 years prior to coming to Australia.

During his time at Mt Wilson he was able to talk - "chew the fat", as Tom would say - with many famous astronomers at a time when astronomy was making rapid advances in the understanding of the universe. Tom has many momentos of his time there. He has a picture when Einstein came to visit the observatory; he has the mechanical calculator that was used to calculate the orbit of Pluto and many of the newly discovered smaller satellites of the giant planets; and he has a painting of probably the rarest observation ever made - a transit of the satellite Triton over the face of Neptune, which he made at the eyepiece of the 100-inch telescope.

He moved on from that when offered a job as Chief Night Assistant at the newly completed 3.9-m Anglo-Australian Telescope. The new (to him) southern skies were the lure, so he and his wife, Mary, came to Coonabarabran in 1976.

At the AAT, Tom helped shape the way support was given at the telescope. Night assistants who knew and understood how the telescope operated were employed, but who also had a daytime role in running the observatory were Tom's preferred model. No "us and them" faction as he had seen happen in some other places. He was also key to making sure that the data recorded at the telescope were properly recorded and archived for posterity. He also helped optimise the telescope observing schedules, kept an eye on the library, fielded calls from the public when they had astronomical (or flying saucer) question. He helped with instruments changes, usually doing the top end changes and telescope balance and anything else that was needed. In the days when we had instruments which needed solid nitrogen (pumped from liquid) he took daily charge of ensuring that it was done properly. On open days, even after he had retired, he would bring along his telescope to show the Sun to the visitors and answer questions.

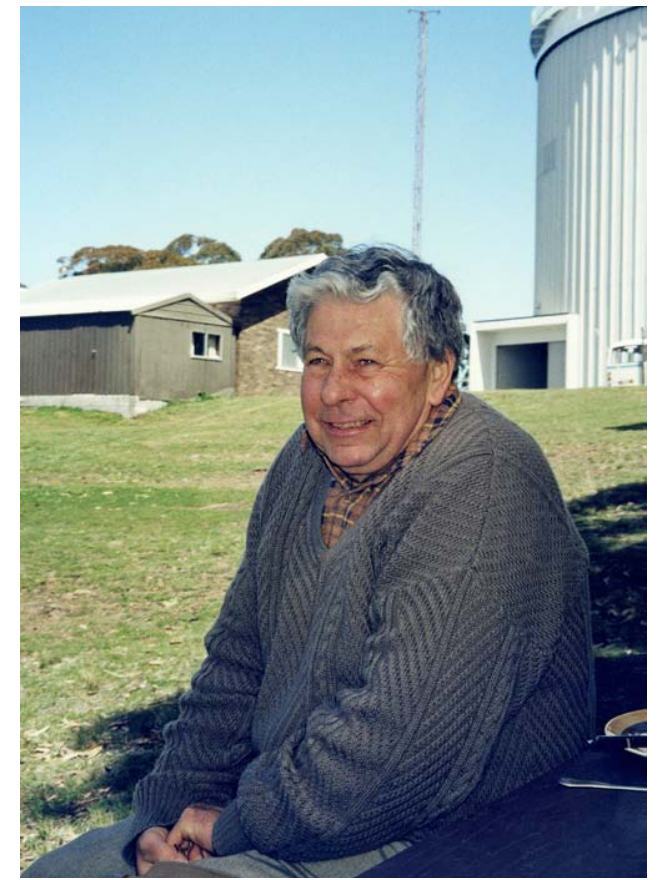

It was to amateur astronomy that Tom really devoted himself. His loves were close to home - the Sun in particular and the planets, and variable stars. None of this cosmology rubbish (science fiction he called it) - you had to understand the stars before you could understand galaxies and the rest of the universe.

Tom was one of the growing band of amateur scientists; making contributions to astronomy through his own personal observations. From 1944 - at the age of about 17 to put it in perspective - he started making drawings of the Sun. He carried on his daily solar drawings for most of the rest of his life - including taking his telescope on holidays so that he didn't miss out. Thousands of drawings in all. I remember him occasionally photocopying drawings to mail back to the observers at Mt Wilson when they had extended cloudy periods - "just so they knew what was happening on the Sun."

At the same time he started his regular solar observations, he started making variable star observations. Long before the days of computer controlled telescopes a very detailed knowledge of the sky was required to locate the star to be observed. The comparison was then done by eye, judging the difference in brightness between the variable and reference field stars. Good observers, like Tom, could do this with almost photometric precision. Exacting work. You then went to the next star in your list. Tom could do perhaps 20 an hour on a good night, although sometimes the going was much slower. He did this for several hours every clear night. For many nights a year. For many years. He amassed well over 160,000 variable star observations in his life.

For his dedication he was given many awards and accolades by his peers, but perhaps the most long-lived recognition was that a celestial body was named after him. On his retirement from the AAT, Rob McNaught named asteroid number 5068 (originally designated as 1990TC) as Cragg.

You might think from this that Tom was only interested in astromomy. Far from it. Amongst his other interests were music, war history and playing war games when he could, plus chess and bridge. He was a founding member of the local astronomical society, as well as a regular player the bridge club and chess clubs in Coonabarabran (when there was one) and in Gunnedah. He played flute with the town orchestra, as well as occasionally accompanying Mary who plays harp.

Tom's funeral at the Coonabarabran Native Grove Cemetry was held on a pleasant sunny day in May and was attended by wide variety of people, many of whom had met him since his retrement and were surprised to learn of his interest in astronomy.

I shared an office with Tom for most of the 16 years he was at the AAT. We shared an interest in astronomy, and I was always fascinated when he would tell stories of the astronomers he'd worked with, the telescopes he'd used and things he'd seen. He was a gentleman, always polite and thinking of others, very generous with both his time and toys (letting me come and use his telescope when he was away) and wonderfully keen and dedicated to astronomy. Astronomy has lost a great observer and recorder of the sky with his passing.

## Letter from Coona

Katrina Harley (AAO)

Well it has been a very busy couple of months at site, with many attending training courses for First Aid and Heights Rescue. In early May the fire team on the mountain conducted burn-offs around the mountain top. Members of the fire team thought it was great experience for the team, and they were very happy with the outcome of the burn.

In February, **Tim Connors** left sunny Coonabarabran for Melbourne. Tim is now working for the Bureau of Meteorology. **Imogen Cosier** joined us in April from Canberra. Imogen is the new IT Support Officer at site, and has settled in well and enjoying Coona life.

In June, the RAVE Workshop was held over two days in Coonabarabran, with reports from many stating it was a very successful and enjoyable meeting. The Acacia Motor Lodge did an excellent job accommodating all of the RAVErs. Many attendees of the workshop travelled back to Vaucluse, to attend a cocktail reception at the residence of the German Consul General.

Earlier last year, **Doug Gray** made a wager with another staff member, (who wants to remain anonymous) that if the unnamed staff member can go without a cigarette for 12 months, then Doug would have to cycle up the mountain. Doug did manage to cycle up the mountain earlier this year and then collapsing in a heap on the ground floor after his accomplishment.

Ending on a high, during the refurbishment of the lifts at the AAT, the OTIS Contractors found a small Black-Headed Python, along with bones of other little critters, in the visitor's lift well. Once the python was rescued from the lift well, it was released back in the bush - one very lucky python.

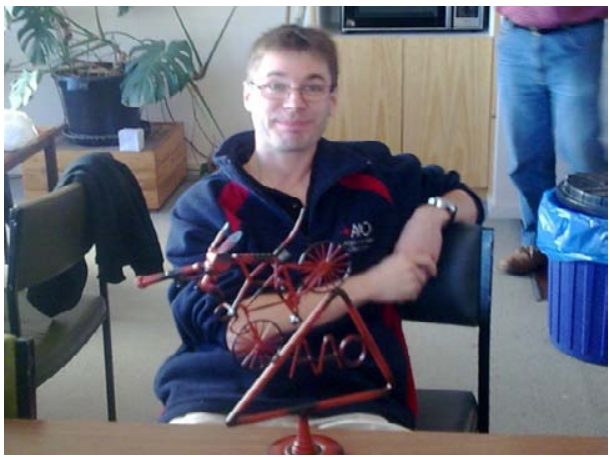

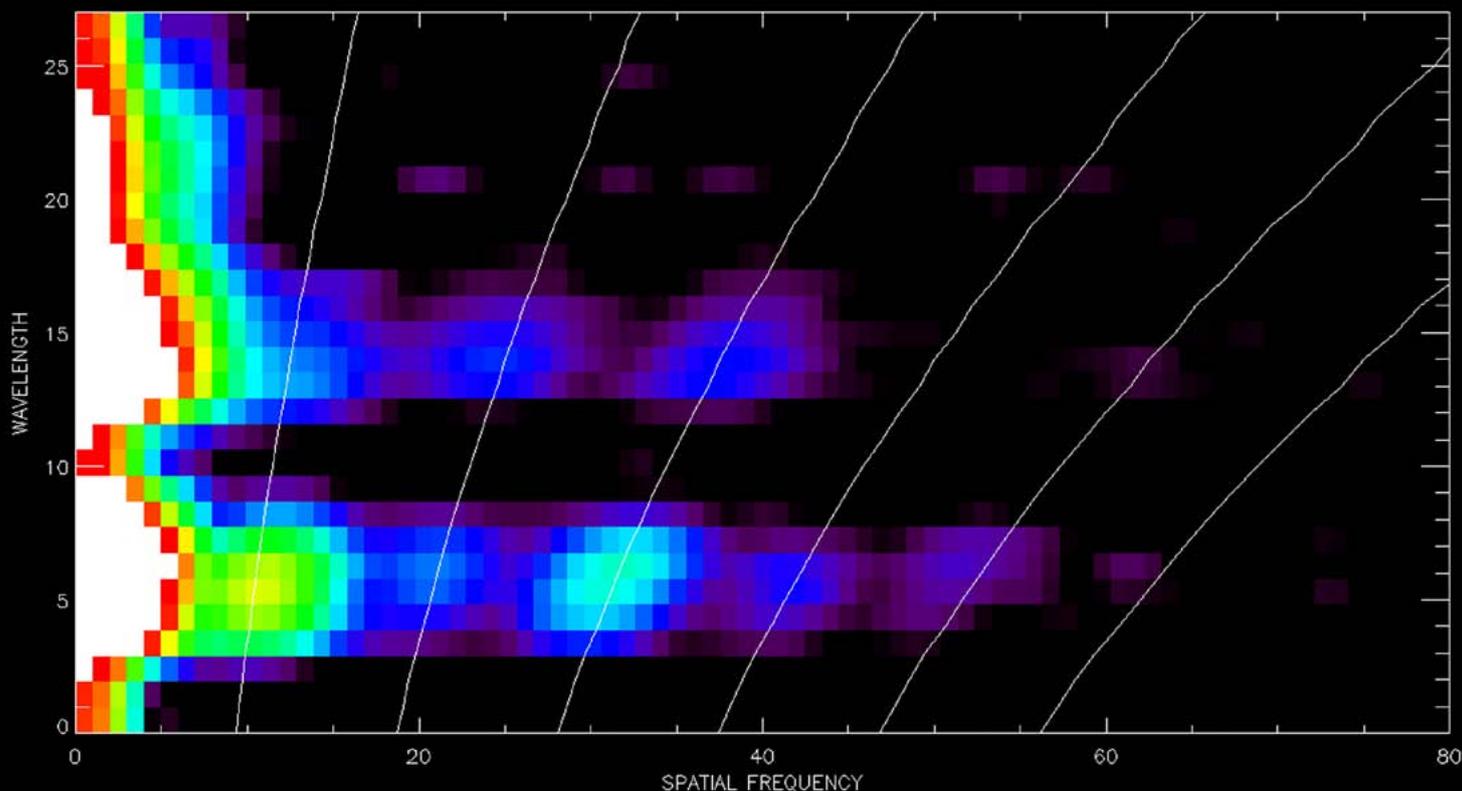

ABOVE: Dragonfly is an optical stellar interferometer taking a revolutionary approach to the design of such instrumentation. This is the power spectrum of the fringes detected from Antares during the Dragonfly commissioning run. Dragonfly is described further on page 7.

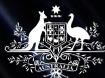
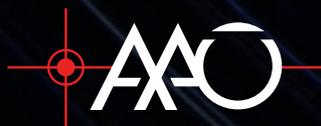